%% file: main.tex
\documentclass[sigconf]{acmart}

\usepackage[T1]{fontenc}
\usepackage{xspace}
\usepackage{color}
\usepackage{xcolor} 
\usepackage{soul} 
\usepackage{enumitem} 
\usepackage{microtype} 
\PassOptionsToPackage{warn}{textcomp}  
\usepackage{textcomp}                  
\usepackage{newtxmath}
\usepackage{tabu}                      
\usepackage{booktabs}                  

\usepackage{svg}
\usepackage{amsmath}
\usepackage{algorithm,algorithmic}

\usepackage{enumitem}

\usepackage{sparklines}

\newcolumntype{L}[1]{>{\raggedright\let\newline\\\arraybackslash\hspace{0pt}}p{#1}} 

\definecolor{brickred}{rgb}{0.8, 0.1, 0.1}
\definecolor{gray}{rgb}{0.8, 0.8, 0.8}
\definecolor{orange}{rgb}{0.85,0.37,0.0}
\definecolor{green}{rgb}{0.0, 0.60, 0.05}
\definecolor{darkviolet}{rgb}{0.58, 0.0, 0.83}
\definecolor{linkColor}{rgb}{0.05,0.5,0.73}

\newcommand{\edit}[1]{{\color{black}#1}}
\newcommand{\secondround}[1]{{\color{black}#1}}

\newcommand{\techname}{ConceptEVA}

\graphicspath{ {./images/} }

\AtBeginDocument{%
  \providecommand\BibTeX{{%
    \normalfont B\kern-0.5em{\scshape i\kern-0.25em b}\kern-0.8em\TeX}}}

\setcopyright{acmcopyright}
\copyrightyear{2023}
\acmYear{2023}
\acmDOI{10.1145/3544548.3581260}

\acmConference[CHI 2023]{CHI Conference on Human Factors in Computing
Systems Proceedings}{April 23-28, 2023}{Hamburg, Germany}
\acmBooktitle{CHI Conference on Human Factors in Computing Systems
Proceedings (CHI 2023), April 23--28, 2023, Hamburg, Germany}
\acmPrice{15.00}
\acmISBN{978-1-4503-XXXX-X/18/06}

\begin{document}

\title{ConceptEVA: Concept-Based Interactive Exploration and Customization of Document Summaries}

\author{Xiaoyu Zhang}
\email{xybzhang@ucdavis.edu}
\affiliation{%
  \institution{Department of Computer Science, University of California, Davis}
   \streetaddress{1 Shields Ave}
   \city{Davis}
   \state{California}
   \country{USA} 
   \postcode{95616}
}

\author{Jianping Kelvin Li}
\email{jpkelvinli@gmail.com}
\affiliation{%
  \institution{Databricks (Work Done at UC Davis)}
   \city{San Jose}
   \state{California}
   \country{USA} 
}

\author{Po-Wei Chi}
\email{pwchi@ucdavis.edu}
\affiliation{%
    \institution{(Work Done at UC Davis)}
   \city{Austin}
   \state{Texas}
  \country{USA}
}

\author{Senthil Chandrasegaran}
\email{r.s.k.chandrasegaran@tudelft.nl}
\affiliation{%
  \institution{Faculty of Industrial Design Engineering, TU Delft}
  \streetaddress{Landbergstraat 15}
  \city{Delft}
  \country{The Netherlands}
}

\author{Kwan-Liu Ma}
\email{klma@ucdavis.edu}
\affiliation{%
  \institution{Department of Computer Science, University of California, Davis}
   \streetaddress{1 Shields Ave}
   \city{Davis}
   \state{California}
   \country{USA} 
   \postcode{95616}
}



\begin{abstract}
With the most advanced natural language processing and artificial intelligence approaches, effective summarization of long and multi-topic documents---such as academic papers---for readers from different domains still remains a challenge.
To address this, we introduce \techname, a mixed-initiative approach to generate, evaluate, and customize summaries for long and multi-topic documents.
\techname\ incorporates a custom multi-task longformer encoder decoder to summarize longer documents.
Interactive visualizations of document concepts as a network reflecting both semantic relatedness and co-occurrence help users focus on concepts of interest.
The user can select these concepts and automatically update the summary to emphasize them.
We present two iterations of \techname\ evaluated through an expert review and a within-subjects study.
We find that participants' satisfaction with customized summaries through \techname\ is higher than their own manually-generated summary, while incorporating critique into the summaries proved challenging.
Based on our findings, we make recommendations for designing summarization systems incorporating mixed-initiative interactions.

\end{abstract}




\maketitle


\section{Introduction}
\input{sections/01-introduction}

\section{Related Work}
\input{sections/02-related_work}

\section{Design Requirements}
\input{sections/03-design_requirements}

\section{Methodology}
\input{sections/04-methodology}



\section{Interface Design}
\input{sections/05-interface_design}

\section{Expert Review of Iteration 1}
\input{sections/06-expert_review}

\section{User Study of Iteration 2}
\input{sections/07-user-study}

\section{Results and Discussion}
\input{sections/08-discussion}

\section{Conclusion}
\input{sections/09-conclusion}

\begin{acks}
\secondround{We thank the participants of our studies and the anonymous reviewers for their feedback and suggestions.
This research is sponsored in part by Bosch Research and the National Science Foundation through grant ITE-2134901.}
\end{acks}

\bibliographystyle{ACM-Reference-Format}

\bibliography{references}

\end{document}

%% file: sections/01-introduction.tex

\begin{figure*}
  \centering
  \includegraphics[width=\linewidth]{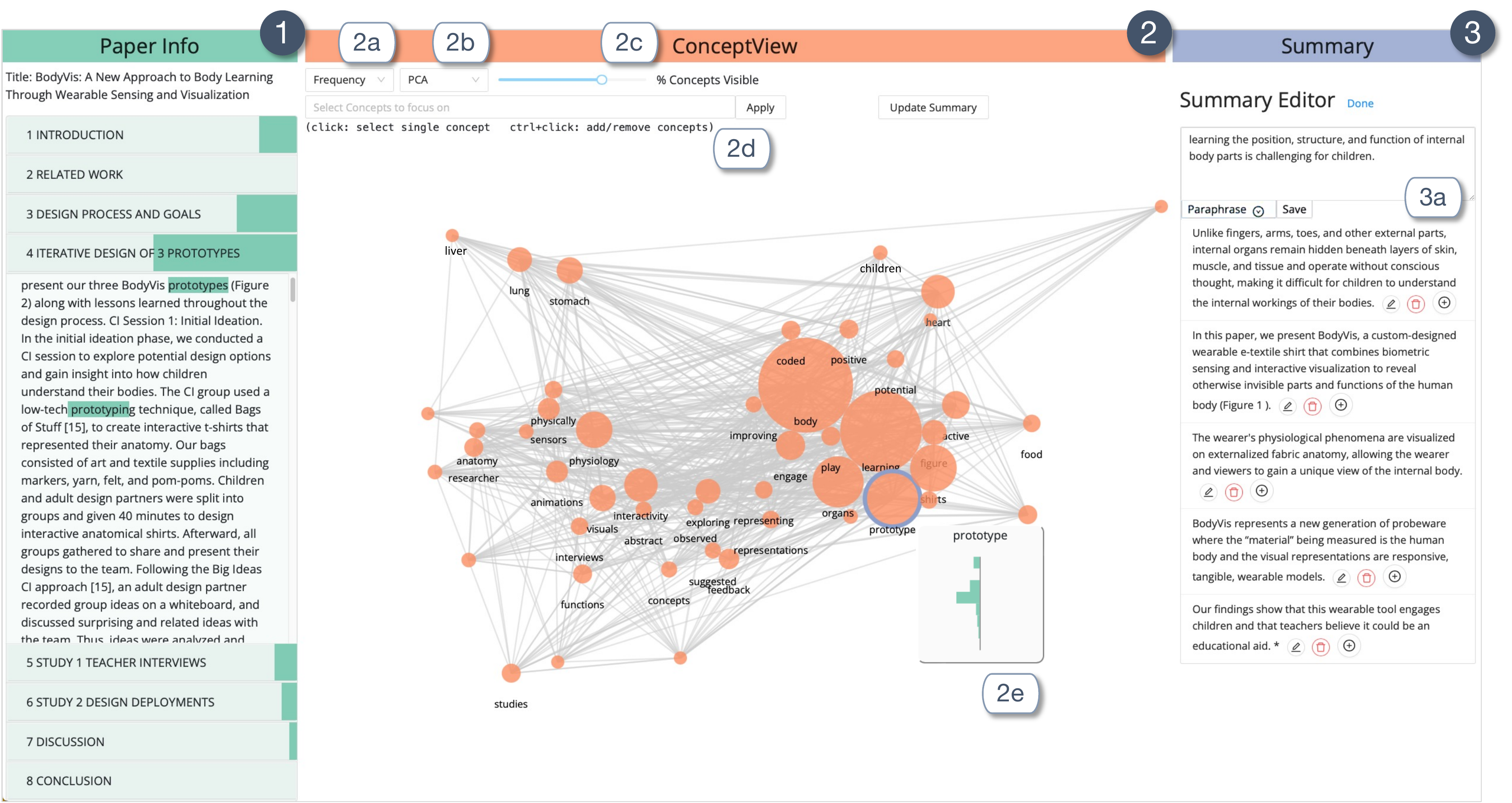}
  \caption{
    The \techname\ Interface shows a multi-disciplinary research paper~\cite{Norooz2015bodyvis} and its auto-generated summary.
    The interface can be separated into three main panels vertically.
    The panel on the left (1) shows a section-wise collapsed view of the research paper, while the panel on the right (3) shows the generated summary in the summary editor.
    At the center is the concept view (2), which displays the concepts extracted from the paper based on a reference ontology.
    The circles represent concepts whose layouts are decided by their text embedding and co-occurrence relationship with other concepts.
    The links indicate whether and how frequently the concepts they connect co-occur in the same sentence.
    A \textit{concept glyph} comparing the concept's appearance distribution across the document and the summary is shown in (2e) upon user request.
    Users can also adjust the concept layout in (2) according to their interests by changing (2a) the information encoded with circle size, (2b) the text embedding projection algorithm, (2c) the percentage of concepts visible, or focusing on the concepts they are interested in (2d). 
    They can also interactively edit (3a), reorder (with drag and drop), or delete any sentences in (3).
    The user can select concepts of interest for the summary generator, which then generates a new summary incorporating the selected concepts.
 }
 \label{fig:interface}
\end{figure*}

The notion of automated text summarization---compression of long text passages into shorter text without losing essential information---has been an open problem since over half a century ago~\cite{Luhn1958automatic}.
The main goals of automated text summarization are to present the salient concepts of a given document in a compact way, and to minimize repetition of the presented ideas or concepts~\cite{El2021automatic}.
Earlier techniques fall under the umbrella of extractive summarization where summaries are generated by extracting terms, phrases, or entire sentences from the source text using statistical techniques~\cite{Gupta2019abstractive}.
With advances in machine learning and specifically sequence-to-sequence language models, abstractive summarization---an approach that generates paraphrased text that still retains concepts from the original text---has gained recent popularity as it mimics summaries created by humans~\cite{Shi2021neural}.

However, significant challenges in abstractive summarization remain, such as the summarization of long, complex, documents that span multiple knowledge domains.
While approaches have been proposed for summarizing domain-specific text~\cite{Lin2000automated} and others for summarizing long documents~\cite{Wu2021recursively}, the challenge remains that there is no one ``ideal'' summary for such long and multi-domain documents.
Automated summarization systems typically do not fare well when the source document spans multiple topics regardless of approach, i.e., extractive~\cite{Gupta2010survey}, abstractive, or hybrid~\cite{El2021automatic}.

Academic papers, especially those in the fields of design or human-computer interaction (HCI) where research tends to be cross-disciplinary, tend to fall under this category of long, multi-topic documents.
For instance, a research article might span the fields of wearable technologies, privacy, and social justice.
A summary of this article that is deemed useful by a researcher in wearable technology would be different from one deemed useful to a researcher in security and privacy.
Yet, both summaries may still be perfectly valid summaries of the article.
This subjectivity means that purely automated, black-box approaches to summary generation will not work.
Instead, a human-in-the-loop approach is needed to allow the user to steer the automated summary generator to interactively generate a summary that is relevant to the user's interests.

To address this challenge, we present \techname, a mixed-initiative system \secondround{for academic document readers and writers to generate, evaluate, and customize automated summaries.}
We build a multi-task Longformer Encoder Decoder (LED)~\cite{Beltagy2020Longformer} from a pretrained LED trained for scientific document summarization by fine-tuning it on two downstream tasks---paraphrasing and semantic sentence embedding---to handle long documents.
This approach uses the notion of attention mechanisms from transformers~\cite{vaswani2017attention} at local levels to reduce memory usage, and at global levels to preserve information fidelity in longer documents.

In addition, \techname\ also supports summary customization for the user by visualizing the concepts---in this scenario, topics explicitly defined in an ontology or knowledge graphs---occurring in the document.
The concepts are identified using a multi-domain ontology~\cite{auer2007dbpedia}, and visualized as a force-directed layout of a graph network using metrics such as concept relatedness and concept co-occurrence in the document.
We introduce a function that we call ``focus-on'', that allows the user to select concept(s) of interest to surface and highlight other concepts related to the selected ones.
The user can identify concepts to focus on and use them to steer the automated summarizer to generate a summary text in which the concepts of interest feature prominently.
The user can also further edit the summary at the concept level by navigating the original document and selecting text to emphasize using the concepts of interest as a filter.
They can also edit it at the sentence level by selecting alternative paraphrasing and sentence ordering.

The design of \techname\ informed by an initial survey of eight research practitioners, and refined through two stages of development and evaluation:

\begin{itemize}[leftmargin=*, noitemsep, topsep=3px]
    \item[] \textit{Iteration 1.} A hierarchical summarization approach with a glyph-based visualization of concepts embedded in a two-dimensional projection, evaluated through an expert review of 3 participants;
    \item[] \textit{Iteration 2.} The final LED approach to summarization described above with concepts visualized as a force-directed layout that preserves both semantic relatedness and concept co-occurrence within the document. This version is evaluated using a within-subjects study of 12 participants using manually-generated summaries as a baseline.
\end{itemize}


Findings from the user study indicate that \techname\ is seen as helpful for participants in examining and verifying ideas, and using specific concepts of interest to explore related concepts and how they are addressed in the source document.
\techname\ was also reported as more useful when the participants evaluated and customized a summary of a document that lay outside their domain of interest, while it was seen as less useful when the participant was knowledgeable about the domain or had a specific idea of what the summary should include.
Using \techname\ for summarization allowed participants to address content-specific aspects of the summary, but inexeperienced participants found it more difficult to incorporate critique such as limitations and implications into the summary.

The chief contribution of this work is \techname, a mixed-initiative system
that integrates interactive visual analysis and NLP techniques for evaluating and customizing long document summaries.
Specifically, we fine tune an LED trained for scientific document summarization for paraphrasing and semantic sentence embedding, identify and visualize concepts from a given academic document using a reference ontology, and provide an interactive visualization system to identify concepts of interest and use them to customize the summary.
We also present insights from a user study on how well users are able to follow summarization guidelines when using \techname.
Finally, we make recommendations for future development and analysis of mixed-initiative summarization systems such as maintaining the user's mental map of the original document by preserving its layout, allowing users to create custom groupings of concepts that will help them add critique to the summary, and minimizing interactive latency for a more fluid interface.

%% file: sections/02-related_work.tex
\techname\ introduces a human-in-the loop, mixed-initiative approach to evaluate and customize document summary generation.
In this section, we review prior work in the domains of summary generation, summary evaluation, and text and document visualization on which we build to create \techname.

\subsection{Summary Evaluation}
\label{sec: related_evaluation}

Summary evaluation techniques can be divided into two main categories: intrinsic~\cite{hariharan2010studies} and extrinsic~\cite{murray2009extrinsic}. 
Intrinsic evaluation methods evaluate a summary based on how well its information matches the information in a reference summary, which is typically human-generated. 
Some examples of intrinsic evaluation of summarization include ROUGE~\cite{lin-2004-rouge} and BERTScore~\cite{zhang2020bertscore}.
Bommasani and Cardie~\cite{bommasani-cardie-2020-intrinsic} propose separate intrinsic scores for compression, topic similarity, abstractivity, redundancy, and semantic coherence.
In contrast, extrinsic evaluation methods evaluate summaries based on their suitability to specific tasks such as following instructions, assessing topic relevance, or answering questions~\cite{Dorr2005methodology, murray2009extrinsic, Hirao2001AnEE}.
In extrinsic approaches, humans subjects are asked to use different summaries to perform a task and uses metrics for their performance---such as completion time and success rate---to evaluate the summaries.


Our work incorporates the principles behind extrinsic summary evaluation methods.
By effectively revealing and comparing the important concepts in a document and its summary, readers can gain confidence in a qualified summary by confirming that it includes all the interested concepts, or see which concepts are missing in a ``poor'' summary.

\subsection{Summary Generation and Customization}
\label{sec: related_customization}
Advances in deep learning and AI has made the automatic generation of good-quality summaries for long document text possible, featured by the success of Transformers~\cite{vaswani2017attention} with its innovative architecture and attention mechanism.
Unsupervised pre-training methods---Masked LM (MLM) and Next Sentence Prediction (NSP)---proposed by Devlin et al.~\cite{devlin2019bert} for their Bidirectional Encoder Representations from Transformers (BERT) enables modeling natural language on a huge corpus, and then fine tuning the model on downstream tasks like summarization.
Inspired by BERT, other researchers~\cite{lewis2019bart, radford2019language, raffel2020exploring, Zhang2020pegasus} propose different pre-training methods and improve the quality of summarization.
For instance, Li et al.~\cite{Li_Zhu_Zhang_Zong_He_2020} propose a multi-task training framework for text summarization that
trains a binary classifier to identify sentence keywords that guides summary generation by mixing encoded sentence and keyword signal using dual attention and co-selective gates.
Wu et al.~\cite{Wu2021recursively} use a top-down approach to recursively summarize long articles like books.
\edit{In our work, we use the Longformer Encoder Decoder (LED)~\cite{Beltagy2020Longformer} for long scientific document summarization, which turns a full attention mechanism---computing relationships between every pair of words in the document---to a local attention mechanism---computing relationships between a more ``local window'' of limited words that precede and succeed any given word.
This has two benefits: faster computation and lower memory usage, which makes it more capable of processing longer documents without a significant drop in the summary quality.}

For the summary customization task, most existing NLP techniques utilizes memory to adjust the auto-regressive language model's output distribution such that the models can retrieve external information given the input prompt. 
Nearest-Neighbour Language Models~\cite{Khandelwal2020Generalization} merge the retrieved information into the output distribution and boost up the language model's perplexity without training.
Borgeaud et al.~\cite{pmlr-v162-borgeaud22a} show that by incorporating a large-scale explicit memory bank, a smaller language model can achieve performance \edit{comparable to  models like GPT-3 with 25 times more parameters, and can update its memory bank without additional training.}
Inspired by these methods, we apply Faiss~\cite{johnson2019billion} to retrieve the k-nearest sentences for each sentence relevant to a chosen concept, and we customize summaries given these sentences as context.

\edit{Besides fully automated approaches, there are also semi-automatic solutions that incorporate humans in the loop.
Post-editing~\cite{moramarco2021preliminary,lai2022exploration} is a common semi-automatic approach for summarizing text, which allows humans to edit AI-generated summaries to ensure accurate and high-quality summarization. Compared to post-editing,~\techname’s approach better exploits human-AI collaboration and iteratively improves the summary by leveraging such collaboration. In contrast to post-editing which only allow human to edit the summary at the end,~\techname supports users to iteratively evaluate and refine the summary by inputting their intention on what should be summarized to the AI models. In \techname’s workflow, users can also edit the AI-generated summary.
But instead of direct manual editing,~\techname leverages AI models to provide aids, such as connections to the concepts, and suggestions for paraphrasing.
}

\subsection{Interactive Visual Analysis for Text Data}
\label{sec: related_visualization}
Our work involves designing interactive visualizations of word embedding and thematic infographics to facilitate summary evaluation and customization.
Visualization of word embeddings~\cite{liu2017visual,smilkov2016embedding,heimerl2018interactive} has been used for supporting text data analysis, such as selecting synonyms, relating concepts, and predicting contexts.
In a different way, thematic visualizations are useful for exploring document and conversational texts.
For instance, ConToVi~\cite{El2016contovi} uses a dust-and-magnet
metaphor~\cite{Soo2005dust} to visualize the placement of conversational
turns (dust) in relation to a set of topics (magnets).
NEREx~\cite{El2017nerex} provides a thematic visualization of
multi-party conversations by extracting and categorizing named entities
from transcripts.
The conversation is then visualized as connected nodes in a network
diagram, allowing a visual, thematic exploration of the conversation.
TalkTraces~\cite{Chandrasegaran2019talktraces} uses a combination of
topic modeling and word embeddings to visualize a meeting's conversation
turns in real time against a planned agenda and the
topics discussed in prior meeting(s).
\edit{
VizByWiki~\cite{lin2018vizbywiki} automatically links contextually  relevant data visualizations retrieved from the internet to enrich new articles.
Kim et al.~\cite{kim2018facilitating} introduced an interactive document reader that automatically references to corresponding tables and/or table cells. 
All these works exploited visualizations to provide contexts or additional information for helping readers to better comprehend text contents. 
}

The application of concept-based clustering is not limited to text analysis: Park
et al.~\cite{Park2021neurocartography} cluster neurons in deep neural
networks based on the concepts they detect in images, and in addition
create a vector space that embeds neurons that detect co-occurring
concepts in close proximity to each other.
Berger et al.~\cite{Berger2016cite2vec} propose cite2vec, a visual
exploration of document collections using a visualization approach that
groups documents based on the context in which they are cited in other
documents, creating a combined document and word embedding.
Closest to our own proposed work is
VitaLITy~\cite{Narechania2021vitality}, an interactive system that aids
academic literature review by providing a mechanism for serendipitously
discovering literature related to a topic or article of interest.
VitaLITy uses a specialized transformer model~\cite{Cohan2020specter} to
aid academic literature recommendations that use additional data such as
citations.
These recommendations are presented via a 2-D projection of the document
collection embeddings generated from the transformer model.
Our work also uses word embeddings to project a view of relevant
concepts onto a 2D space, but is different from VitaLITy in the purpose:
our focus is on interactively exploring the concept focus of a generated
summary as well as generating summaries that emphasize concepts of
interest within an academic publication.

In our work, we use visualization of word embeddings to provide overviews of all the important concepts in a document and identify which concepts are missing in the summary for evaluation.
Thematic infographics is used in the visualization of word embedding to show the details and occurrences of a concept in both the document and summary for comparison. 


%% file: sections/03-design_requirements.tex
\label{sec:requirements}
In order to better understand the different requirements and motivations when summarizing an academic article, we conducted a preliminary survey of 8 higher education professionals: one professor, 4 associate professors, and 3 assistant professors (7 male, 1 female, all between 25--44 years of age).
The survey covered open-ended questions concerning how they motivated and guided students' paper summaries, how they evaluated such summaries, and what they consider to be a good summary and why.

Based on the experts' responses, we grouped their remarks and suggestions under three categories: \textbf{\textit{process}}, representing approaches they use or suggest students to follow in order to summarize an academic document; \textbf{\textit{content}}, representing what should be included in the summary; \textbf{\textit{requirements}}, representing attributes that make for a ``good'' summary.
Each remark or statement below is suffixed with a count showing the number of experts who shared the corresponding opinion.

\begin{itemize} [leftmargin=*, noitemsep, topsep=3px]
    \item \textsc{Process:} Approaches to follow when summarizing.
        
        \begin{itemize}[leftmargin=1.5em]
            \item  Prioritize referring to the abstract, conclusion, introduction, and title (7 experts).
            \item  Use the abstract \& introduction as a ``backbone'' for the summary (1 expert).
            \item  Familiarize oneself with background and context, then identify strengths \& weaknesses (1 expert).
            \item  Find parts of the paper relevant to one's context or interest and focus on them (1 expert).
        \end{itemize}
        
    \item \textsc{Content:} What the summary should include.
        \begin{itemize}[leftmargin=1.5em]
            \item An Explanation of what the paper is about and what its contributions are (5 experts).
            \edit{
            \item The major ideas of the proposed solution and its difference from prior work (3 experts).
            \item The results generated by the solution, and how they address the problem/research question (3 experts).
            }
            \item The problem addressed by the paper and the research questions it answers (2 experts).
            \item An outline of existing approaches to address the research question or problem, their advantages and limitations, and the challenges (2 experts).
            \item The advantages/disadvantages of the solution and the strengths/ weaknesses of the paper (2 experts).
        \end{itemize}
        
    \item \textsc{Requirements}
        \begin{itemize}[leftmargin=1.5em]
            \edit{
            \item The summary should have an indication that the summarizer has not simply paraphrased the paper but also thought about and understood the underlying ideas (3 experts). 
            \item The summary should show reflection on the ideas and discuss implications for practice/research. (3 experts)} 
            \item The summary should include a figure if possible (2 experts). 
            \item The summary should have a clear structure \& emphasis (2 experts)
            \item The summary should be specific and provide details, paraphrasing where necessary and quoting from the paper where necessary (1 expert). 
        \end{itemize}
\end{itemize}


While the above responses are relevant for manual summarization, we also examined existing approaches of evaluating automated summarization techniques, such as fluency, saliency, novelty, and coherence~\cite{Tan2017abstractive}. 
Saliency is an especially complex issue as saliency of a given summary may vary across readers depending on each reader's background and research focus.
Based on the responses and on prior work on automated summarization, we synthesized the following requirements that we prioritize for mixed-initiative approaches that help the user evaluate and customize summaries of scientific articles:

\begin{enumerate}[leftmargin=*, noitemsep, topsep=3px]
     \item[\textbf{R1}] \textbf{Accuracy Evaluation:} The technique should help the user efficiently verify whether a summary
     accurately reflects the content of the original document based on the criteria established by the user (see \edit{R4}: Flexibility below).
     This requirement is synthesized from participant responses categorized under \textit{``criteria''} and \textit{``structure''}.
    
     \item[\textbf{R2}] \textbf{Provenance Evaluation:}
     The technique should show direct or indirect contributors to a summary to help the user verify whether the summary reflects the structure and key components of the original document.
     This includes the parts of the original document---a research article in this case---that contribute to the summary.
     It also includes external references (see R3: Contextualizations) that influence parts of the summary.
     This requirement is synthesized from responses under \textit{``topics''}, \textit{``structure''}, and \textit{``strategies''}.
     
     \item[\textbf{R3}] \textbf{Contextualization:} The technique should be able to provide some context in which the work presented in the paper exits.
    Such a context includes the contribution of the work, as well as the significance of the work, its strengths, weaknesses and so on. This can include information presented within the paper itself but should not be restricted to it.
    This requirement is based on the participant responses under \textit{``criteria''}.
     
    \item[\textbf{R4}] \textbf{Flexibility:} The technique should be flexible enough to change the summaries based on the priority of the user. 
    For instance, the summary may focus on the relevance of the paper to a concept of interest to the user.
    Alternatively, the summary may also be one that examines the paper's contributions, approach, and methods---or any combination thereof.
    The requirement is based on participant responses under \textit{``topics''} and \textit{``strategies'}.

\end{enumerate}

%% file: sections/04-methodology.tex
In \techname, we support summary evaluation and customization by empowering the 
exploratory visual analysis (EVA) with multiple natural language processing (NLP) techniques. In this section, we first introduce the data processing and visual analysis framework of \techname, then describe the major NLP techniques backing the functionalities.

\subsection{Framework Overview}

\begin{figure*}
  \centering
  \includegraphics[width=\textwidth]{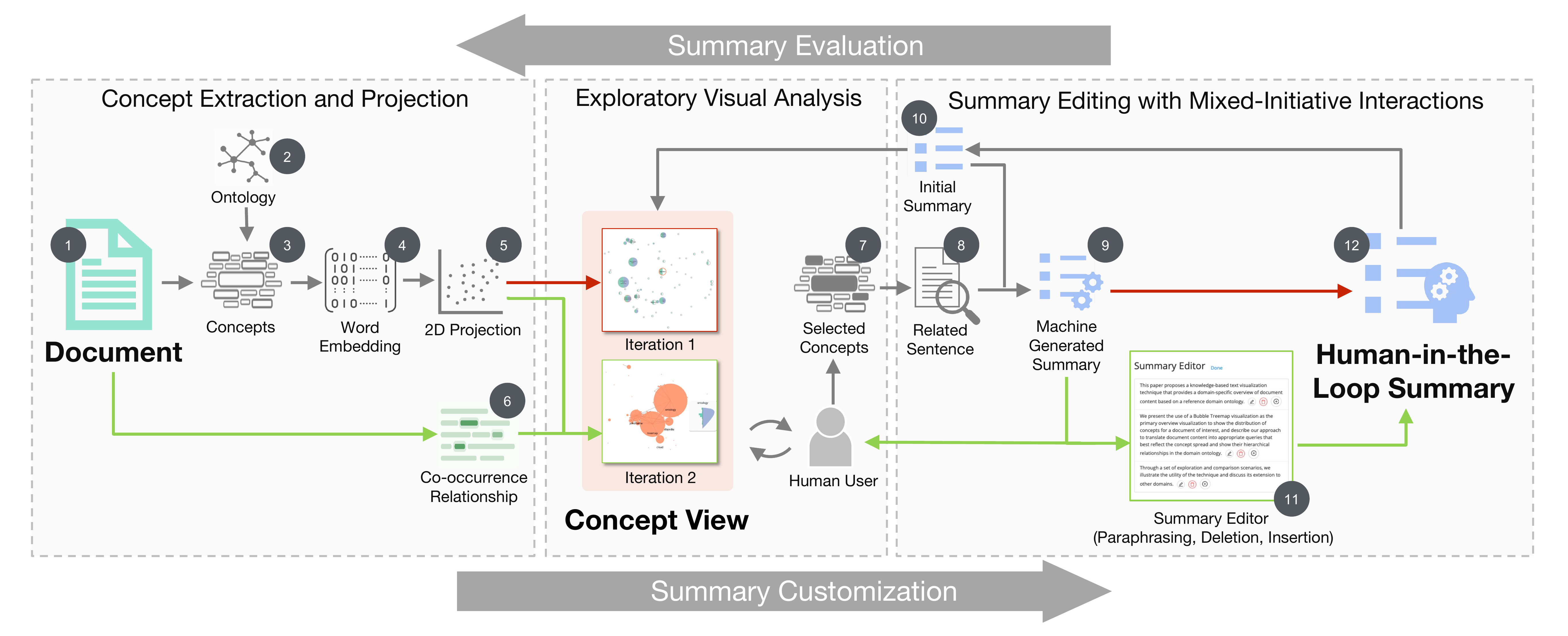}
  \caption{
    The framework of \techname. The core idea is to bridge the document and the summary with a concept view. In iteration 1, the concept view shows an embedding-based layout that allows users to select concepts to include in the machine-generated customized summary (red boxes \& arrows). In iteration 2, the concept view also includes the co-occurrence information in a force-directed layout, and a summary editor with mixed-initiative interactions is added (green boxes \& arrows). In both iterations, the user can repeat the human-in-the-loop summary customization for multiple rounds till they are satisfied with the result.} 
 \label{fig:framework}
\end{figure*}

\techname\ leverages knowledge graphs, NLP, and EVA techniques to facilitate summary evaluation and customization for academic document readers.
We bridge the original document and the summary with a concept view visualizing all of the concepts identified from the document.
As shown in Fig.~\ref{fig:framework}, we start by extracting concepts from an academic document according to a reference ontology, converting them into text embeddings and projecting them onto a two-dimensional space (Sec.~\ref{sec: method_knowledge_base}).
After that, we present the semantic and contextual information of the concepts in an interactive visual interface that supports flexible concept exploration and customized concept(s) prioritizing (Sec.~\ref{sec: method_EVA}). 
Finally, we provide an interactive summary editor to facilitate dedicated refinement of a new version of the summary we generated according to the user-specified concepts of interest (Sec.~\ref{sec: method_summary_edit}).
In this way, we help the users evaluate the quality of an  AI-generated summary and see how well it addresses the readers' focus of interest in the paper, as well as support them customizing the summary to alter their specific requirements if the automated one is not satisfied enough.



\subsubsection{Concept Extraction and Projection}
\label{sec: method_knowledge_base}

In order to effectively extract the key concepts from a large body of texts, knowledge graphs, such as DBpedia~\cite{auer2007dbpedia}, Freebase~\cite{bollacker2008freebase}, and Wikitology~\cite{rahm2001survey}, can be used to look up established concepts in specific domains.
We use DBpedia-Spotlight~\cite{10.1145/2063518.2063519} to extract concepts and rank their importance by term frequency.
We then visually highlight concepts to show which ones are included or missed in the AI-generated or customized summary.
To vectorize these concepts, \techname\ leverages text embeddings to represent concepts, sentences, and descriptions of the concepts as high-dimensional vectors.
Two-dimensional projections of these ``concept vectors'' are computed using dimensionality reduction techniques, such as PCA\cite{10.1162/089976699300016728}, t-SNE~\cite{JMLR:v9:vandermaaten08a}, or UMAP\cite{mcinnes2020umap}.
Semantically similar concepts are placed closer together in the projections, while different concepts are placed farther apart.



\subsubsection{Exploratory Visual Analysis}
\label{sec: method_EVA}

To allow readers to explore and reason about the concepts, \techname\ provides interactive visualizations to help trace these concepts back to the source document text as well as to the generated summary.
A visual representation (see Sec.~\ref{sec:visualization-interaction} for details) is designed to show the importance of the concepts and help the user compare their occurrences in the document text and the summary.
Readers can not only use \techname's interactive visual interface to explore and understand each concept, but also select concepts that are relevant to their research interests.
The selected concepts are used to recompute the importance and relevance of each concept in the high-dimensional embedding and recreate the projection, allowing the readers to ``steer'' the exploration.

\subsubsection{Summary Editing with Mixed-Initiative Interactions}
\label{sec: method_summary_edit}

While generating a good summary that can satisfy the user's needs and interests cannot solely rely on NLP techniques, \techname\ provides a set of mixed-initiative interactions for quickly customizing and editing an AI-generated summary.
From the user interface, users can easily select which concepts in the document are important or match their interests.
If the generated summary did not provide enough context or description of these concepts, the user can indicate where in the summary that they want to add a sentence about a particular concept, then \techname\ will immediately generate a list of sentences that describe that concept for the user to choose.
In addition, \techname\ allows users to paraphrase any of the sentences based on its NLP models.

\subsection{Natural Language Processing: Multi-Task Longformer Encoder Decoder}
\label{sec: method_NLP2}

\edit{
As shown in Fig.~\ref{fig:framework}, \techname\ uses several NLP techniques at various stages of summary generation and customization.
At the center of these techniques is a multi-task Longformer Encoder Decoder (LED)~\cite{Beltagy2020Longformer} that we develop for iteration 2. We describe in this section the motivation to use LED and its functions at specific stages in summary generation and customization.
}

\edit{In the first iteration of \techname, we developed a a hierarchical summarization method with BERT Extractive Summarizer~\cite{bert-extractive-summarizer} and a Pegasus abstractive summarizer~\cite{Zhang2020pegasus} for summary generation and customization of long documents (please refer to supplementary materials for more details).
However, this approach could easily incur high interaction latency caused by sentence clustering and iterative summarization of long documents.}
To alleviate these issues, we develop for the second iteration a multi-task Longformer Encoder Decoder (LED)~\cite{Beltagy2020Longformer}, capable of processing longer documents.
In addition, we take advantage of weight sharing, \edit{i.e., every task shares weights on the common parts of the network's memory,} thus optimizing the time and space efficiency of \techname\ and speeding up the system's responses to human input.

Our multi-task LED is employed in \techname\ for four functionalities: scientific document summarization, paraphrasing, semantic text encoding, and summary customization (see Fig. \ref{fig:nlp_functionalities}).
We describe \edit{these functionalities} below.

\begin{figure*}
  \centering
  \includegraphics[width=\linewidth]{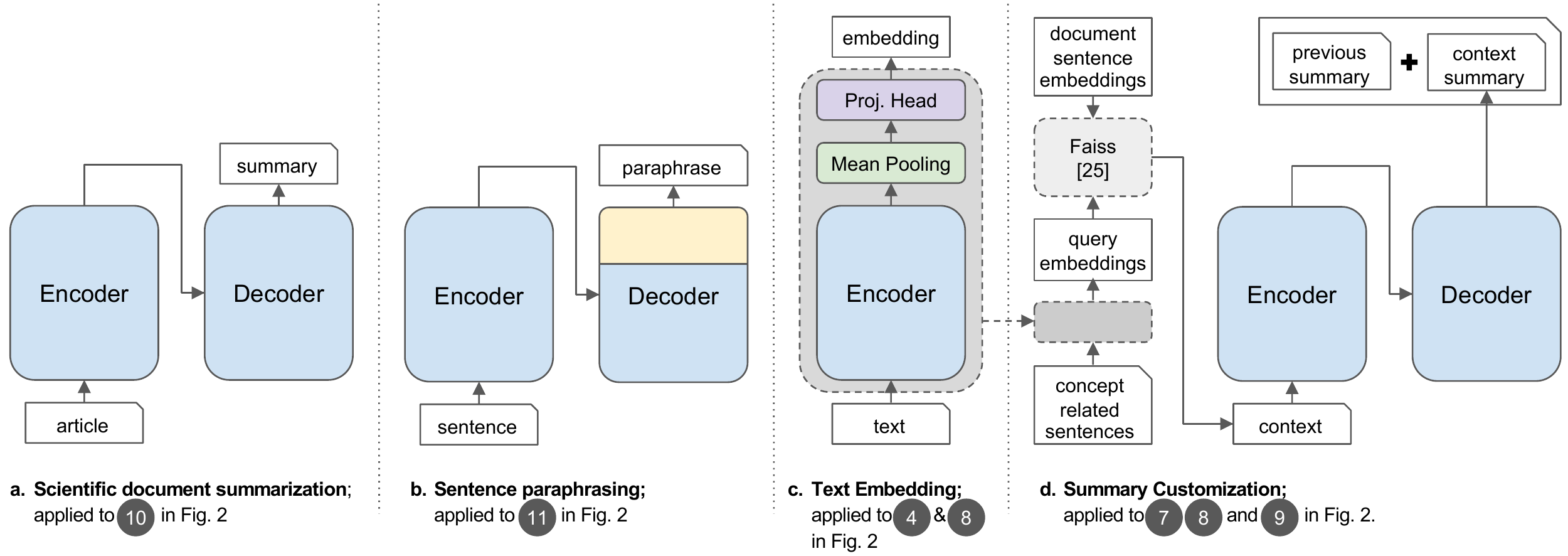}
  \caption{
  \edit{\techname\ uses a multi-task LED model~\cite{Beltagy2020Longformer} to help generate, evaluate, and customize summaries.
  Specifically, LED performs four functions, shown above as subfigures a--d with detailed explanations in Sec.~\ref{sec: method_NLP2}.
  The text below each subfigure indicates the corresponding function in Fig.~\ref{fig:framework} for which the model is used.
  In each subfigure, the blue rounded boxes represent the weights from the LED trained for summarizing scientific papers and shared across all tasks.
  The yellow, purple, and green rounded boxes represent fine-tuned layers for downstream tasks.
  The functions are: (a)~\textbf{Scientific document summarization:} The LED's training data, local self-attention mechanism, and high memory complexity make it suitable to summarize academic papers.  
  (b) \textbf{Sentence paraphrasing:} We fine tune the last two decoder layers (shown in yellow) with a set of ``paraphrasing datasets''---datasets that contain multiple paraphrases of a given set of sentences. This helps in generating alternative sentences for a given sentence when editing a summary. 
  (c) \textbf{Text embedding:} To generate the concept layout (see Fig.~\ref{fig:interface}-2) and fetch relavant context for summary customization, we compute
  text embeddings---vector representations of concepts or sentences in a high-dimensional space.
  This is done by adding a mean pooling layer (green) and a projection head (purple) to the encoder and fine-tuning it (see Sec.~\ref{sec: method_NLP2} for details).
  (d) \textbf{Summary customization}. Pre-computed embeddings of every sentence in the source document are queried using vector representations---retrieved from the text embedding shown in (c)---of user-selected concepts. Nearest sentences are appended to provide `context' for the selected concepts, and then summarised. The resulting summarized sentences are appended to the existing summary (see details in Sec.~\ref{sec: method_NLP2}).
  }}
  \label{fig:nlp_functionalities}
\end{figure*}

\textbf{Scientific Document Summarization:}
\edit{The LED model was trained on the ArXiv dataset of scientific papers~\cite{clement2019arxiv}.}
Due to its local self-attention mechanism, the memory complexity of LED grows linearly, making it capable of handling up to 16384 tokens, which is \edit{typically long enough for handling academic papers}. 
\edit{These factors render the LED suitable for generating summaries of academic papers.
These automatically-generated summaries (see item `10' in Fig.~\ref{fig:framework})} act as a starting point for users to evaluate and customize upon according to their interests.

\textbf{Text Paraphrasing:}
\edit{One of the functions in \techname's mixed-initiative interactions is the ability to paraphrase text, or specifically, generate alternative summaries for a selected sentence.
To achieve this capability, we fine tune the pre-trained model on relatively small datasets with small learning rates.
We ``freeze all the layers'' of the model, i.e., we keep all model weights the same during training except for the last two decoder layers.
The decoder takes a sequence of tokens as the input and generates the next token based on its weights.
We train these two decoder layers on a dataset that contains 147,883 sentence pairs, with each pair containing two alternative paraphrases of one sentence (Fig.~\ref{fig:nlp_functionalities}b).
We build this dataset by merging three other datasets: PAWS~\cite{paws2019naacl}, MRPC from GLUE~\cite{wang2018glue}, and TaPaCo~\cite{scherrer-2020-tapaco}.
Once fine-tuned, this model is capable of taking as input one sentence and providing a paraphrased sentence as an output.
In item `11' in Fig.~\ref{fig:framework}, this model is accessed via the summary editor when the user opts for automated paraphrasing of a selected sentence.
} 


\textbf{Text Embedding:}
\edit{To generate the concept layout (see Fig.~\ref{fig:interface}-2) and fetch relevant context for summary customization, text embeddings---representing the relationships between concepts or sentences in a high-dimensional space---need to be computed.
To compute sentence embeddings, we follow the siamese network architecture from SentenceBERT~\cite{reimers-2019-sentence-bert}, an approach to generate sentence embeddings, i.e., vector representations of sentences that preserve semantic relationships.
We add a `mean pooling layer'---a function that averages the embeddings of input tokens---and a `projection head'---a function that computes a high-dimensional space that captures semantic similarities between all sentences---on the LED's encoder (Fig.~\ref{fig:nlp_functionalities}c).
We then fine-tune the encoder for learning meaningful sentence embeddings by freezing all layers of the encoder and only training on the projection head.
For the training data, we once again follow SentenceBERT: we combine the SNLI \cite{snli:emnlp2015} and MultiNLI \cite{N18-1101} datasets, and format each data sample as a triplet of an `anchor sentence', a `positive sentence', and a `negative sentence'.
The training involves fine-tuning the embedding such that in each triplet, the positive sentence ends up closer to the anchor sentence than the negative sentence.
We also follow data augmentation approaches (detailed in the supplementary materials) inspired by those followed in SentenceBERT~\cite{reimers-2019-sentence-bert}.
The resulting model is used in two main functions of \techname: generation of the ``concept view'' (see Fig.~\ref{fig:framework}), the ``focus-on'' function (detailed in Sec.~\ref{sec: vis_evaluation}), and subsequent summary customization (see items `7' and `8' in Fig.~\ref{fig:framework}).
}

\textbf{Summary Customization:}
\edit{\techname\ customizes a generated summary by updating it to include concepts of interest selected by the user.
To achieve this, we pre-compute embeddings for every sentence in the source document.
When a user selects a concept or concepts of interest, we retrieve corresponding text embeddings using the model described in the previous paragraph.
We then use these embeddings as `queries' to search for sentences in the pre-computed embeddings that are closest to the query vectors (see Fig.~\ref{fig:nlp_functionalities}).
We apply Faiss~\cite{johnson2019billion}---a similarity search library of dense vectors in large scale---to implement this approach.
The nearest sentences are concatenated in the order of their appearance in the original document and included in the input to the summarizer as `context' for the selected concepts.
The resulting, newly-summarized sentences are then appended into the previously-generated summary.
In this form of summary customization, new concepts add to the existing summary but do not result in the erasure of parts of the existing summmary.
The summary editor provides the option for the user to manually delete the sentences.}


%% file: sections/05-interface_design.tex
\label{sec:visualization-interaction}
The \techname\ interface (Fig.~\ref{fig:interface}) consists of three main panels: a document view on the left (with green header \& accents) that collapses into a section-wise overview, a summary view (blue header \& accents) on the right displaying the generated summary and associated metadata, and a central concept view (orange header \& accents) showing the relative dominance and associations between the concepts found in the document.
Additional controls for visualizing and filtering the concepts are also provided on top of the concept view.
The interface design has gone through two iterations of development, incorporating feedback and insights from the expert review (Sec.~\ref{sec: expert_review}).
We detail the visualization and interaction design choices of \edit{the final version of the system}
and the underlying rationale in this section. 

\subsection{Concept View: Document-Summary Relations}
\label{sec:vis_metaphor}
In the concept view, we provide an overview of the document-summary relation from the perspective of concepts.
We represent each of the concepts occurring in the documents as a node---a ``concept circle''---the size of which shows the dominance of the concept in the source document.
User-specific metrics of dominance, such as ``frequency'' and ``tf-idf'' are available for the user to choose.

\begin{figure}
  \centering
  \includegraphics[width=\columnwidth]{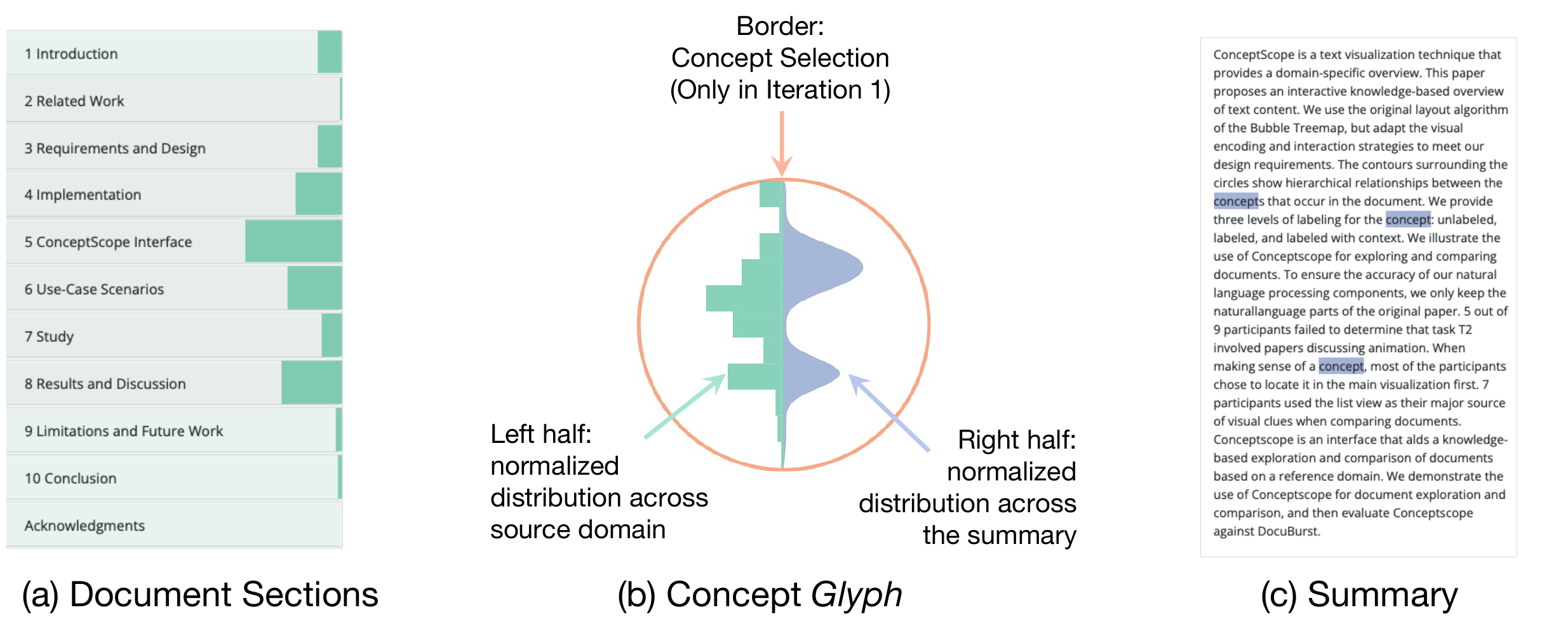}
  \caption{
    The \textit{concept glyph} extends the concept circle to support the in-place comparison of concept distribution between the document and the summary. This glyph is shown for all dominant concepts in iteration 1 and in a floating tooltip upon request in iteration 2.
 }
 \label{fig:concept_glyph}
\end{figure}

To convey information about the structure of the document and of the summary (\textbf{R2}), we incorporate the user's orientation to the interface---the document on the left and summary on the right---into the concept view to represent concepts that are present in the document and concepts present in both the document and the summary.
We design the \textit{concept glyph}---a pair of histograms representing the distribution of the concept across the source document and the summary respectively (see Fig.~\ref{fig:concept_glyph}).
The histograms are oriented vertically and share a common axis.
This way, the histogram on the left indicates the source document and the curved line on the right (histogram smoothed with a kernel density estimation) represents the summary.
The number of bins on the histogram on the left matches the number of sections in the source document, while the right one maps to the number of sentences in the summary.
\edit{For instance, the concept ``prototype'' is missing in the summary shown in Fig. \ref{fig:interface} because the right half of the glyph is missing.}
To further reinforce this connection between the histogram and the document view, 
we create an echo of the histogram overlaid on top of the section headers (Fig.~\ref{fig:concept_glyph}-a).
This allows the user to identify the sections of the document in which the concept is most dominant, and examine those sections closely if needed.

When determining the two-dimensional(2D) layout of these concepts on the concept view and the amount of information to reveal for each of them, \edit{we started with an embedding-based layout in iteration 1 where the \textit{concept glyph} of every concept were displayed and distributed according to the text embedding (Fig. ~\ref{fig:concept_view_comparision}-a, see supplementary materials for details.)}
\begin{figure*}
  \centering
  \includegraphics[width=.9\linewidth]{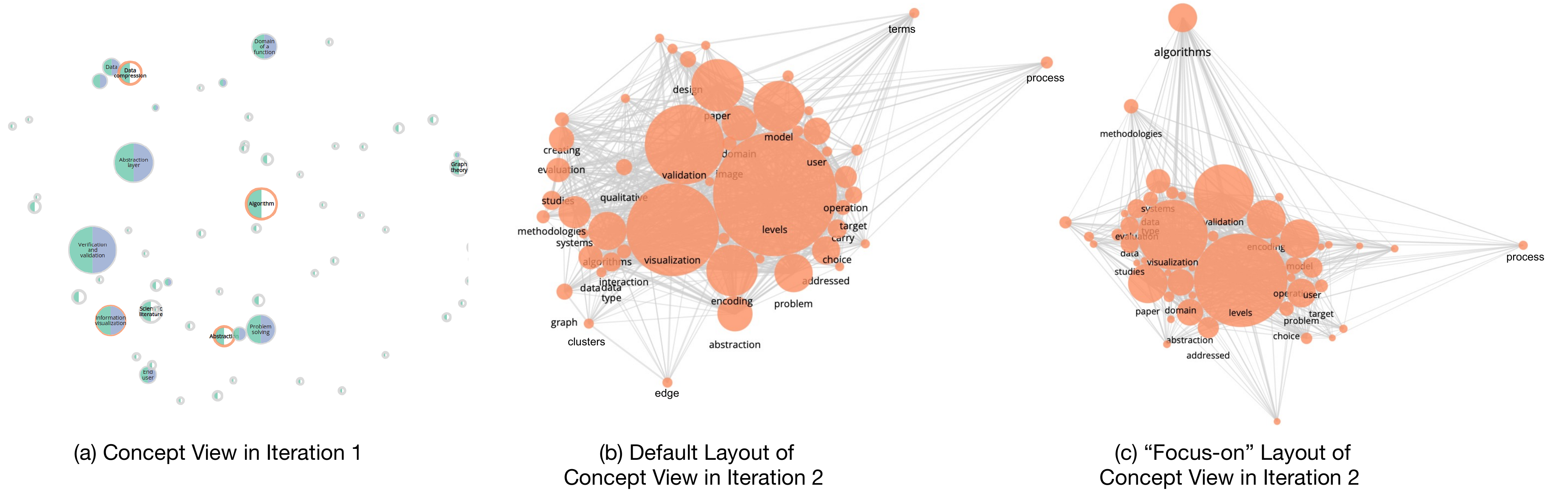}
  \caption{
    A comparison of the embedding-based layout in iteration 1 and the context-augmented latyout in iteration 2 for the concept view. All three figures show 80\% of the concepts from the paper~\cite{munzner2009nested}. The circle size represents frequency (The size scale and ontology query parameters are slightly different between iteration 1 and 2). The "focus-on" layout in (c) focuses on the concept ``algorithm''.
 }
 \label{fig:concept_view_comparision}
\end{figure*}
\edit{While this layout was designed to help the user efficiently} compare the occurrence of concepts in the original document against those in the summary, the expert review results (Sec.~\ref{sec: expert_review}) indicated that showing such a comparison for all the concepts in one visualization was too overwhelming to the users.
\edit{To reduce such perception load, we shifted to} a more intuitive visualization design in \edit{iteration 2}
where the visual representation of the concepts \edit{were simplified} to solid circles (Fig.~\ref{fig:concept_view_comparision}-b) in a force-directed layout.
The coordinates of these circles are initialized by a 2D projection of the concepts' semantic word embedding and adjusted by links representing the co-occurrence relationship of two concepts in the same sentence \edit{per experts' request for more co-occurrence information support}.
In this way, we created a context-augmented layout with the coordinates of each concept influenced by both its semantic meaning and its co-occurrence relationship with the other concepts in the specific academic document (\textbf{R3}).
\edit{For instance, Fig. ~\ref{fig:interface}-2 shows that the concepts ``organ'' and ``prototype'' are semantically remote but co-occur frequently in~\cite{Norooz2015bodyvis}, while aligns with the fact of this document.}
Our context-augmented layout could capture such document-specific concept co-locations and adapt the initial text embedding in concept view to reflect the document context.
\edit{To efficiently support the user to evaluate the summary quality from the perspective of concept appearance, we move the \textit{concept glyph} with detailed document-summary information for each concept to a tooltip which can be triggered by hovering in iteration 2.}
This provides an effective overview and detail-on-demand exploration of the concepts in a document using interactive visual analysis.

\subsection{Summary Evaluation}
\label{sec: vis_evaluation}

\edit{To facilitate the users to get an intuition about concepts from the document that are included in the summary compared the the concepts excluded from the summary (\textbf{R1}), we designed the \textit{concept glyphs} (Fig. \ref{fig:interface}-2e) as described in~Sec.\ref{sec:vis_metaphor}.}
Users can quickly filter out all but the ``important'' concepts, and then compare their distribution and context in the document and in the summary using the \textit{concept glyphs} and the linked view to the document on the left (\textbf{R3}).
To cater to user-specific analysis requirements (\textbf{R4}), we allow users to (1) choose the criteria (frequency or tf-idf) by which concepts should be considered ``important'' (\autoref{fig:interface}-2a), (2) choose the dimensionality reduction method (PCA, tSNE, or UMAP) to project the concepts (\autoref{fig:interface}-2b and 3), and filter them to only show the top K percent of concepts based on ConceptEVA's importance metric (\autoref{fig:interface}-2c).


\edit{Inspired by the experts' attempt to locate concepts with the ``focus-on'' function and their significant interest in it, we enhanced the ``focus-on'' function in iteration 2} to allow the user to switch perspectives and evaluate how well the current version of the summary addresses their specific areas of interest (\autoref{fig:concept_view_comparision}-c).
When the user triggers the ``focus-on'' function, they will be able to select from full list of the concepts sorted by their appearance frequency in the original document  (\autoref{fig:interface}-2d). 
Users can select one or multiple concepts based on their research interests and trigger a corresponding update of the concept view layout.
The concept they choose to focus on will ``float to the top'' of the concept view, i.e., move to the top of the view, and the rest of the concept will ``sink'' to the bottom, with semantically or contextual-wise more relevant concepts pulled higher towards the top and less relevant concepts pushed lower towards the bottom.
Meanwhile, the horizontal layout remains to reflect the concepts semantic and contextual distance determined by the user-chosen projection method.
For instance, the layout in Fig.~\ref{fig:concept_view_comparision}-c was focused on the concept ``algorithm''.
We can see the related concepts including ``methodologies'', ``validation'', and ``systems'' are also pulled upwards.
Meanwhile, the layout of the remaining concepts Fig.~\ref{fig:concept_view_comparision}-b is locally maintained, continuing to reflect their semantic and contextual closeness in the document.
This will further facilitate the concept selection and inform the customization task described in Sec.~\ref{sec: vis_customization}.


\subsection{Summary Customization}
\label{sec: vis_customization}

\edit{Reflecting on the requirements we collect for a ``good'' summary (Sec~\ref{sec:requirements}),} we approach summary customization in two ways: at a concept level, we see summary customization as determining what concepts are included when generating the summary, while at a structural level, we see it as inserting, reordering, and rewriting content.
Users can achieve the concept-level summarization by selecting a group of concepts from the concept view to prioritize for the next version of the summary.
Based on user selection, the summarizer extracts relevant sentences from the document as described in Sec.~\ref{sec: method_NLP2} and inputs them to the summarization pipeline for a customized summary that better addresses the concepts of interest.

The AI-generated summarization approach focuses more on the content than the flow of the summary, and was seen in the expert review as compromising the logical and narrative connection from one sentence to the next (see Sec.~\ref{sec: expert_review} for details).
To address these concerns about the summary quality,
we extended the interactions supported in \techname~with an interactive summary editor to facilitate better human-AI collaboration in iteration 2. 
With the AI-generated summary as a starting point, the summary editor (\autoref{fig:interface}-3) helps users iteratively customize or extend the summary (\textbf{R1} \& \textbf{R2}) by:
(1) choosing from a list of candidate sentences for all user-selected concepts categorized by concept name, and inserting them into the summary,
(2) updating a particular sentence in the summary with automatically paraphrased sentences generated with the paraphrasing model in Sec.~\ref{sec: method_NLP2} (\autoref{fig:interface}-3a), and
(3) interactively editing, reordering, or deleting any sentences. 
In this way, a human-in-the-loop summary will be generated as the final output of the summary customization process in which user knowledge and judgments are effectively cooperated with the NLP techniques described in Section~\ref{sec: method_NLP2}.


%% file: sections/06-expert_review.tex
\label{sec: expert_review}

Iteration 1 of \techname\ was evaluated through expert review with three participants (2 male, 1 female).
\edit{Given our prototype was backed with a NLP model more suited for scientific document analysis, we invited three experts with Ph.D.\ degrees in computer science with InfoVis as their research focus.}
Participant details are listed below, with years of experience in reading/reviewing academic papers included in parentheses.


\begin{enumerate}[itemindent=0.5em]
    \item[\textsc{E1:}] software engineer (5--10 years).
    \item[\textsc{E2:}] senior applied scientist and former academic (10--20 years).
    \item[\textsc{E3:}] data scientist (5 to 10 years).
\end{enumerate}

The review was conducted online via a video conference setting. 
Participants were first introduced to \techname's functions and features and given trial tasks with a test dataset to familiarize them with the interface.

Participants then used \techname\ to finish two open-ended tasks while following a concurrent think-aloud protocol:
(1) verify the auto-generated summary for a given document, and
(2) generate a customized summary according to a set of requirements provided to them. 
Since the participants were experienced researchers in infovis, we also collected their feedback and recommendations on the system as suggestions to incorporate into iteration 2.
Iteration 1 was received positively in general, especially idea of evaluating a document summary by examining the concepts (E1, E2, E3), context and support views to compare the document and the summary (E2, E3), but the quality of the generated summary was not considered sufficient (E1, E2, E3).
Specific feedback is listed as follows:

\begin{itemize} [leftmargin=*, noitemsep, topsep=3px]
    \item \begin{sloppypar}\textbf{\textit{Concept Extraction \& Separation:}} Concept identification through fuzzy matching between document terms and the reference ontology sometimes produced results that the experts (E1, E2, E3) found confusing.
    Iteration 1's implementation of the ``focus-on'' interaction was also not deemed helpful likely due to the issues concerning the fuzzy matching (E1, E2, E3), \edit{though all experts expressed considerable interest and pointed out potential ways for improvement.
    E1 and E2 also expressed that they expected a better-functioned ``focus-on'' tool with more intuitive interaction.
    E3 also suggested providing concept searching functions, showing the frequency of the concepts, and sorting the searching list accordingly.}\end{sloppypar}
    
    \item \textbf{\textit{Information support:}} The visual representation of the concepts and the way they supported the comparison of the summary against the document was deemed helpful (E2, E3).
    Showing co-occurrence information of concepts was recommended (E1, E2, E3).
    
    \item \textbf{\textit{Summary quality and presentation:}} An initial paragraph-like summary shown to E1 \& E2 was deemed to not have a logical flow, while a bullet-point format change with E3 was received well.
    However, E3 was uncertain on how well they could ``trust'' the summary if it were of an unfamiliar paper, and recommended showing additional information to increase the user's confidence in the summary.
    
\end{itemize}

%% file: sections/07-user-study.tex
\label{sec:user_study}
Lessons learned from the expert review helped focus the redesign of \techname\ and focus its evaluation through tasks that reflect how a researcher may approach summarizing an academic paper.
Specifically, we decided to focus our study on whether and how a participant is able to generate a summary of a paper with which they are familiar using \techname\ such that the summary is relevant to their research interests.

While comparing the use of \techname\ with an existing summarization tool would be ideal, to our knowledge there is no existing summarization tool designed for research documents.
We thus chose human-generated summaries by each participant as the baseline for that participant.
While this means there is no ``standard'' baseline across all participants, this approach gives us better ecological validity as each participant would generate a summary that is relevant to their own interests and research contexts.
Therefore, the current baseline for researchers would be to generate a summary by themselves---unaided by other tools.
This would serve two purposes.
Firstly, by generating their own summary manually, they gain familiarity with the document and are able to use \techname\ as a tool to refresh their memory, navigate the concepts relevant to the document, and be able to compare the summary they generate using \techname\ against their own manually-generated summary.
Secondly, the process serves to emphasize our idea that \techname\ is \textit{not} intended as a replacement for reading the document; it is intended to augment the way the document is explored.

This necessitated a \edit{study with a within-subjects component}
where each participant first generated a summary manually before attempting the same task on \techname.
For the same reason, there was no counterbalancing: asking all participants to perform the manual summarization task first allowed us to ensure they were familiar with the document before they used \techname.
It also allowed participants to critically examine the extent to which they could create a summary that was relevant to their own interest in the document.
\edit{We used two test papers~\cite{Norooz2015bodyvis, zhang2021conceptscope}, one for six participants in this study.}

\subsection{Participants}
We recruited 12 participants (4 female, 8 male, aged 25--44 years), comprising 10 Ph.D. students, 1 university faculty, and 1 research engineer from a technology company. 
Seven participants reported they had been actively reading academic papers for 5-10 years, and the remaining five reported less than 5 years.
And 10 participants reported they had written a summary/abstract/short description for an academic paper more than 10 times before the study, and the remaining two did it for 3-10 times.
Two of the 12 participants reported themselves as native English speakers.

\subsection{Experimental Setup}

We conducted the study remotely considering the varied geographical locations of the participants and a safety measures surrounding the uncertain conditions of COVID-19.
Instructions for the offline study task \textbf{T1} were shared with participants no less than 12 hours before the online study session began.
For the online study session, the participants were asked to access \techname from a remote server and participate in the study with their own machine and external devices.
Six participants used the Chrome browser with the Windows operating system, four used Chrome with MacOS, and the remaining two used the Safari browser with MacOS for the tasks.
The setup, tasks, and durations were decided based on a pilot study with three participants: one native and two non-native English speakers.

We asked the participants to follow the ``think aloud'' protocol and \edit{audio- and video-recorded them during the task.}
Each participant received a $\$10$ Amazon gift card as a compensation for their participation.

\subsection{Summarization Guidelines}
\label{sec:summ-guidelines}

Based on findings from our survey of research practitioners explained in Sec.~\ref{sec:requirements}, we constructed a set of guidelines for participants to follow when generating a summary manually or using \techname.
The guidelines were presented in the form of the following list of questions that participants could try and answer in their summary.

\begin{enumerate} [leftmargin=*, noitemsep, topsep=3px]
    \item[\textbf{G1}] \textbf{\textit{Content.}} What is the paper about? What are the contributions?
    \item[\textbf{G2}] \textbf{\textit{Approach.}} If the paper addresses a problem, how does it do it?
    \item[\textbf{G3}] \textbf{\textit{Comparison.}} If the paper addresses a problem, how does its approach compare to existing approaches to address the same problem?
    \item[\textbf{G4}] \textbf{\textit{Insights.}} What insights does the paper offer from its analysis or evaluation of the approach?
    \item[\textbf{G5}] \textbf{\textit{Critique.}} What are the strengths and weaknesses of the approach?
    \item[\textbf{G6}] \textbf{\textit{Implications.}} What are the implications of the work to your own interests and/or research?
\end{enumerate}

We made it clear to participants that they were free to choose some, all, or even none of the guidelines below when generating the summary.
In the procedure below, we would ask the participants which of the guidelines they followed for each summarization process: manual and using \techname.

\subsection{Procedure}

Each participant was provided with a research paper a few days in advance of the scheduled session with the study moderator, along with the guidelines listed in Sec.\ref{sec:summ-guidelines}.
Each participant was then assigned the following tasks:

\begin{itemize} [leftmargin=2em, topsep=3px]
    \item[\textbf{T1:}] \textbf{Manual summarization.}
        \begin{itemize} [leftmargin=-1em, noitemsep]
        
            \item We asked participants to read the paper and manually generate a summary between a minimum of 100 and a maximum of 150 words reflecting what they found interesting in the paper.
            This summary was to be sent to the moderator in advance of their scheduled session.
            This represents the baseline for each participant, indicating the summary they would generate without \techname. 
            It also ensures that participants read the paper before the start of the study.
            
            \item After their summary was received, participants were also asked to fill in a survey relating to their background and demographics. They were also asked to respond on a 7-point Likert scale (one for each guideline in Sec.~\ref{sec:summ-guidelines}) the extent to which they followed the guideline.
            
            \item Participants were also asked to report on their experience of the summarization task on the NASA TLX scale~\cite{Hart1988development}.
            
        \end{itemize}
        
    \item[\textbf{T2:}] \textbf{Automated summarization.}
        \begin{itemize} [leftmargin=-1em, noitemsep]
        
            \item Participants were shown the automated summary generated without human intervention and asked to read through it.
            
        \end{itemize}
        
    \item[\textbf{T3:}] \textbf{Human-in-the-loop summarization.} 
        \begin{itemize} [leftmargin=-1em, noitemsep]
        
            \item Participants were introduced to the \techname\ interface and allowed to explore it through mini-tasks that reflected the process they would follow in their main task. This training/exploration session used a paper different from the one used for their tasks.
            
            \item Participants were then instructed to generate a summary of the same paper as in T1, following the same prompts and guidelines, but this time using \techname\ to explore and focus on concepts of interest and choosing relevant concepts to steer the summary generated. 
            Throughout this exploration participants were instructed to follow a concurrent think-aloud protocol where they verbalized their thinking during their exploration.
            
            \item At the end of this process, they responded to a 7-point Likert scale (same as in T1) showing the extent to which they followed each guideline from Sec.~\ref{sec:summ-guidelines}.
            
            \item Participants reported on their experience of the summarization task on the NASA TLX scale.
            
        \end{itemize}
    \item[\textbf{T4:}] \textbf{Rating all summaries.}
        \begin{itemize} [leftmargin=-1em, noitemsep]
        
            \item Participants finally rated on a 7-point Likert scale their satisfaction with (a) their manually-generated summary from T1, (b) \techname's automated summary with no human intervention from T2, and (c) the summary they generated in T3 using \techname\ by focusing on concepts of interest.
            They were allowed to re-read all three summaries before reporting on their satisfaction.
            \secondround{The reason behind choosing ``satisfaction'' as a metric and for having participants rating their own summaries as opposed to others' summaries are related. Recall that the reason behind proposing \techname\ was that different readers of the same research article may emphasize different aspects when generating a summary of the paper. A participant with their own concepts of interest in a given paper would have takeaways that are influenced by these interests, which would in turn be reflected in their summary of the paper. We deemed that it would be less insightful for them to evaluate a summary generated by a different participant with different interests and takeaways. Instead, having the participant examine the summaries they have themselves created through three approaches could potentially reveal more insights into how well the human-in-the-loop approach has worked, as each participant can examine all summaries through the lens of their interest in the paper. For the same reason, ``satisfaction'' as a measure along with participant responses explaining the reasoning behind the rating allows us a way to understand what aspects of human-in-the-loop summarization are valuable for participants, albeit at the expense of specific insights more objective measures may provide.}
            
        \end{itemize}
\end{itemize}

\edit{The study did not focus on speed or quality of task performance, but on participants' own satisfaction with their experience and outcome.
Thus task times were not restricted, and we did not track the time participants spent on Task 1, only their self-reported experience in writing the summary as described above.
Participants in general spent between 60 and 90 minutes on tasks T2--T4.}

\begin{figure*}[t]
  \centering
  \includegraphics[width=0.75\linewidth]{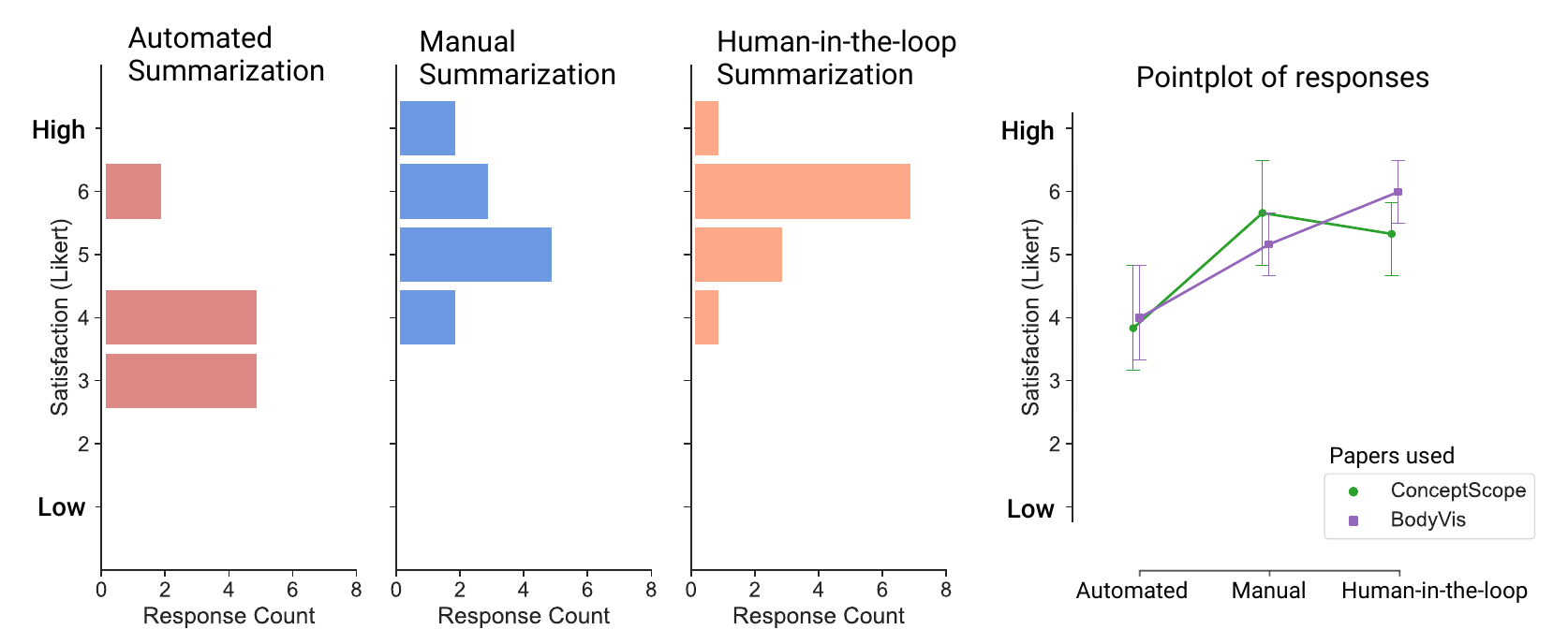}

  \caption{
    \edit{Distribution of participant responses on a 7-point Likert scale showing their level of satisfaction with the summaries from the automated approach in task \textbf{T2}, manual approach from task \textbf{T1}, and the human-in-the-loop approach in task \textbf{T3} created with \techname.
    The three charts on the left show the distribution as response counts for each summarization approach.
    The right chart shows average values for each approach for the two papers used in the study, ConceptScope~\cite{zhang2021conceptscope} and BodyVis~\cite{Norooz2015bodyvis}, with error bars indicating 95\% confidence intervals.}
 }
 \label{fig:quality_rating}
\end{figure*}


%% file: sections/08-discussion.tex
\label{sec: Results and Discussion}
\noindent

\subsection{Summary Satisfaction}
\label{subsec:summary_quality}

As mentioned in Sec.~\ref{sec:user_study}, we used each participant's manually-generated summary (\textbf{T1}) as a unique baseline for that participant.
Ten of the 12 participants rated the automated summary (task \textbf{T2}) \textit{lower} than the baseline, 8 out of 12 participants rated the summary generated using \techname's human-in-the-loop approach (task \textbf{T3}) \textit{higher} than the baseline (Fig.~\ref{fig:quality_rating}).
\edit{Recall that two papers were used in the study---6 participants summarized ConceptScope~\cite{zhang2021conceptscope} and 6 summarized BodyVis~\cite{Norooz2015bodyvis}.
Fig.~\ref{fig:quality_rating} also includes a pointplot showing average ratings split across both papers.
While the small participant pool makes it difficult to state with sufficient confidence whether participant satisfaction with the human-in-the-loop summarization using \techname\ is equivalent to their satisfaction with their own manually-generated summary, Fig.~\ref{fig:quality_rating} suggests such an equivalence.
In addition, a chi-squared test of independence showed a significant association between summarization approach and summary satisfaction rating, $\chi^2\left(8\right) = 23.5, p < 0.01$.
On the other hand, a chi-squared test of independence showed no significant association between the paper used and summary satisfaction rating, $\chi^2\left(4\right) = 0.87, p = 0.93$.
This indicates that the differences seen in Fig.~\ref{fig:quality_rating} are more likely to be due to the summarization approach rather than the paper used in the task.
}

Participants who gave a higher rating for the human-in-the-loop approach reported being able to locate and focus on concepts more efficiently (P4, P6, P9), and on the content of the summary itself (P7).
P7 observed that ``\textit{the contribution of this paper, was also well described in the \emph{(human-in-the-loop generated)} summary.}''
Participants who preferred the manual version of their summary to the human-in-the loop approach (P1, P3, P11) explained that they had their own idea of a summary that they wanted the generated version to reflect.
For instance, P11 wanted the summary to focus on the paper methodology, and deleted all sentences from the automated summary, directing the system to pull new sentences from the paper focusing on ``visualization'', ``concept'', and ``ontology''.
They proceeded to edit these new sentences based on their recall of the document and even manually wrote some text from scratch.
These participants also reported a lower level of trust in the AI component of \techname\ through the study.

Participants' level of trust in the generated summary also appeared to be influenced by their confidence in their knowledge of the domains addressed in the paper.
For instance, BodyVis~\cite{Norooz2015bodyvis}, one of the papers used in the study, covers domains like participatory design, physiological sensing, and tangible learning, which the participants were relatively unfamiliar with.
Their response to the summary generated by \techname\ was more positive.
P4 reflected that \textit{``in terms of ... describing the (BodyVis) system, maybe the one generated by \techname\ is kind of better... In the manually generated summary, although I put my focus there, I didn't do a good job like mentioning it. I don't think if I mentioned it.''}
P10 noted the automated summary addressed some of their own omissions: \textit{``In my manual summary. I actually skipped some details, like I didn't really mention ... the feedback from children and the teachers (about BodyVis).''}
In contrast, for a topic they were knowledgeable in, participants seemed to prefer their own interpretations and emphases, as P1 states: \textit{``For the papers, if I already know that area, I have a certain expectation of what I need to look at. Then I would still prefer to write the summary by myself.''}  



In terms of the process, all participants reported being able to follow guidelines G1 (content) and G2 (approach) i.e., they rated themselves above 4 on a 7-point Likert scale.
Six out of 12 participants reported being able to follow G4 (insights) and G6 (implications) as shown in Fig.~\ref{fig:experience_rating}.
Participant ratings on being able to follow G3 (comparison) and G5 (critique) were skewed heavily toward the lower end of the scale.
Participants P4 and P8 found it the most difficult to address these two guidelines, and they had a common approach: they attempted to find concepts related to ``limitations'' or ``cons'' to see the weaknesses reported in the paper itself and found this approach difficult to critique the paper and compare it with existing work.
A low chance of success is expected with this approach as it is difficult to critique a paper by only examining the paper without a general sense of the related work.
A summary that features such critique is difficult to automate as it would need knowledge as well as critical thinking about related work.



\subsection{Summarization Experience}
\label{subsec:experience}

\begin{figure*}
  \centering
  \includegraphics[width=0.8\linewidth]{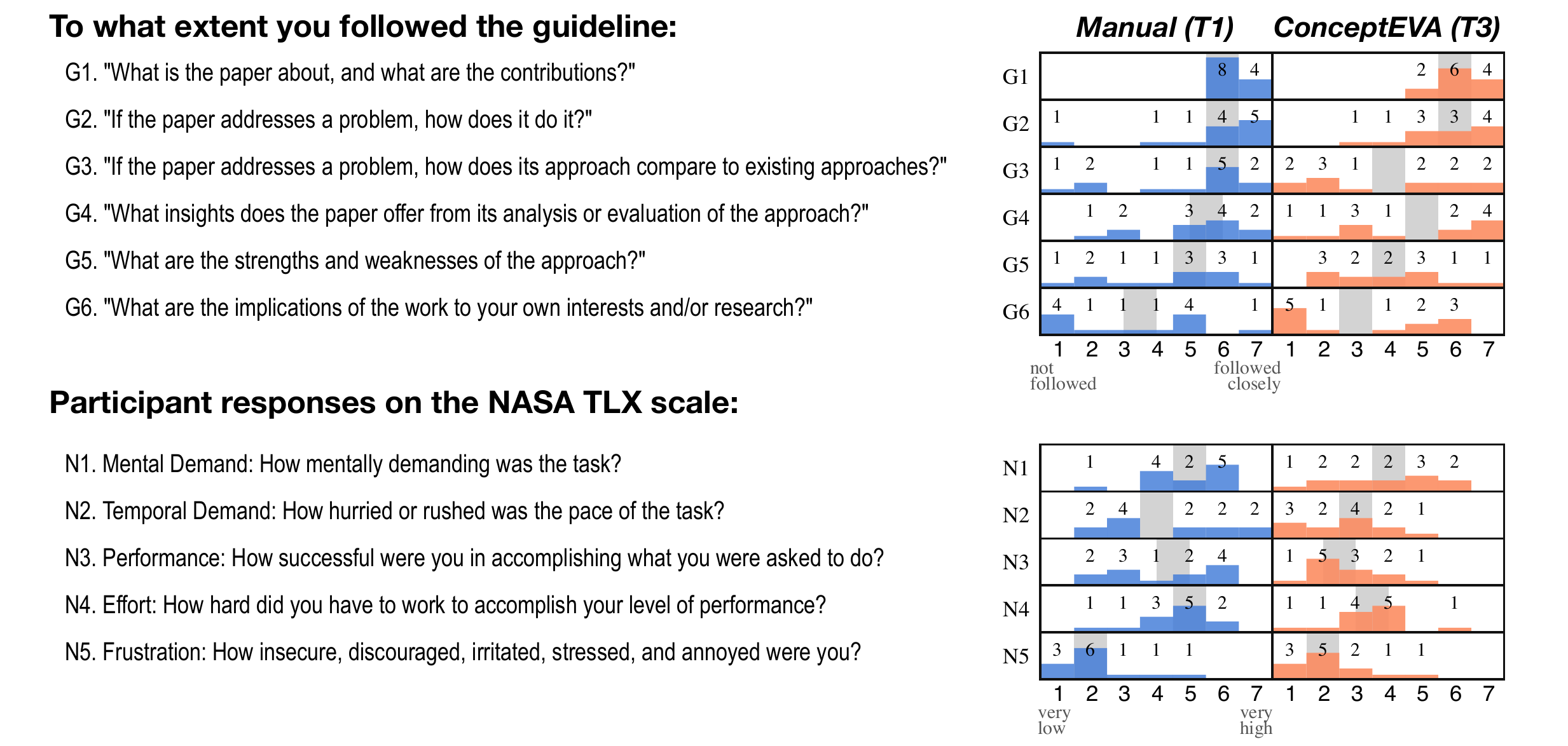}
  \caption{Ratings for manually-generated summary in T1 and human-in-the-loop summary in T3. Median ratings are in gray.}
 \label{fig:experience_rating}
\end{figure*}

When responding to the NASA TLX scale (see Fig.~\ref{fig:experience_rating}) and rating their summarization experience, participants described the experience of using \techname\ as ``\textit{helpful}''(P1, P3, P7, P9, P11), ``\textit{useful}''(P1, P4, P6, P8, P9), ``\textit{amusing}'' (P5) and ``\textit{enjoyable}'' (P5).
Eight participants reported that the concept view provided useful information such as the importance,  appearance frequency, and co-occurrences of concepts.
P4 and P6 also reported finding the focus-on function helpful to explore relationships with less dominant concepts. 
``\textit{sometimes a concept is kind of minor...sheltered by those big circles...but by lifting it up you can see all the relation to other concepts. you can also like, and identify it directly.}'' (P4).

The glyphs from the earlier iteration that were redesigned to be revealed only on detailed inspection were also deemed helpful by 6 participants, indicating perhaps that the glyph in isolation was helpful but several together were distracting.
Since \techname was implemented for the browser, we also observed participants incorporating built-in browser functionalities such as search, translation (for bilingual/multilingual participants), and grammar checkers.

Participants also expressed their frustration when they ``\textit{can't find anything useful about the word they identified}''(P3) or ``\textit{lose the full picture of the paper}''(P8).
Identifying relevant concepts is a function of the ontology, and a balance between the specificity of domain ontologies and the breadth of a general ontology such as DBpedia.
On the other hand, issues related to identifying strengths and weaknesses of the work that may not be explicitly stated in the paper---echoing participant experiences described in Sec.~\ref{subsec:summary_quality}---may be possible to address by additional visualization of document affect and sentiment~\cite{Kucher2018state}.

\subsection{Influence Factors on User Experience}
\label{subsec:insights}

When conducting tasks \textbf{T3} and \textbf{T4}, we observed four dominant factors that appeared to influence participants' use and preferences of certain functionalities in \techname.
\edit{Please refer to Table 3 in the supplementary materials for a detailed report of participants' behavior patterns when performing the user study tasks.}

\begin{itemize} [leftmargin=*, noitemsep, topsep=3px]
    \item \textit{Academic reading experience and skill influences exploration.} Participants such as senior PhD students and faculty/researchers preferred to read the original text of the paper. P5, a graduate student with 5--10 years of experience reading academic papers, said they preferred to read the original text of the paper, but also said that the concept view \textit{``is actually really good with the way my brain is... I just think of words, and then it (the focus-on function) has the words I want. This kind of maps with my thinking, which is very amusing.''} 
    In contrast, participants with either less academic experience or from a different domain 
    \edit{found direct text reading difficult.
    For example, P7 thought the paper reading process was``\textit{very overwhelming}'' while P8 reported that they ``\textit{don't have the full picture of the paper in this way}''}.
    They preferred to use the visualizations---the projection view or the Focus-on too---to get a high-level overview, and then ``\textit{grab information based on the concept that I'm giving}''(P11). 
    While this is part of the intention behind designing the visualizations (esp. \textbf{R3}), a longitudinal study may be needed to explore how \techname\ may be used as a way to scaffold students' ability to read and understand academic text.
    Note that P5, P7, P8, and P11 are all graduate students, but P5 identifies as a native English speaker while the others do not.
    While this may not be the reason for the difference, it brings up the issue of reading skill, a factor that was not evaluated in the study.

    
    \item \textit{Academic writing experience influences summarization.}
    An extension of the above observation means that participants' academic writing experience would influence how they used \techname\ to summarize text.
    P5 found the workflow afforded by \techname\ useful, and that it was ``\textit{doing most of the work for me}'', such as ``\textit{constructing sentences I would put in my paper, or something letting me take what either my problem is or what I'm thinking about looking at the paper, and like merging these things together}''.
    They also appreciated ``\textit{the freedom of allowing more editing}'' in the summary editing panel (\textbf{R4}), and used it to directly edit the summary sentences.
    P10 reported finding it useful to ``\textit{pull out the related sentences categorized by each of the concepts you selected}'' (\textbf{R3}).
    Other experienced participants like P11 reported that \techname\ ``\textit{doesn't encode (sic) their standard of generating the summary}.''
    Note that P11 is also the participant who heavily edited the generated summary (Sec.~\ref{subsec:summary_quality}).
    
    
    \item \textit{Domain familiarity influences use of \techname.}
    Participants' reflections indicated that their knowledge of the domain covered in the document would influence how they would use \techname.
    P1 mentioned that ``\textit{if I'm reading a machine learning paper or deep learning one that I'm not quite familiar with (the domain)}'', they would prefer to use the concept view to ``\textit{understand what kind of concepts they (the paper) have}'' and would like to see definitions of the concept in \techname.  
    On the other hand, for documents in their own domain, they said they would \textit{``have a certain expectation of what I need to look at. Then I would still prefer to write the summary by myself.}''
    This was also seen in P8's approach in the study: they were unfamiliar with the paper they were asked to read and requested more information support as they did not have ``\textit{a general picture of the paper.}''
    
    
    \item \textit{Mental map of document influences use of visual interface.}
    Eleven of the 12 participants reported being happy with the visual interface for the summary customization task.
    While distributing their time to the three panels in \techname in different ways, 11 out of the 12 participants embraced the visual interface for the summary customization task in our study.
    Participants preferred different aspects of the interface depending on the way they approached ideas in the paper.
    P5, quoted earlier in this section, stated how the concept view layout mirrored the way they think.
    P11, on the other hand, preferred the ``paper info'' panel to the concept view \textit{``because I can see I know where it (the concept) is (in the paper).''}
    They even chose to search the concept directly in the PDF version of the paper after briefly exploring the Focus-on function in the concept view panel, explaining that ``\textit{it's quite a huge number of information ... it's a little bit hard to draw the connection between the information inside the original paper and the (concept view) exploration panel. That's why I just ignore the exploration panel.''}
    Others found the paper info panel disorienting as it provided a view of the paper that was different from the PDF layout they had initially read, stating, \textit{``I don't have, like the mental map of the original pdf. It's gone''} (P5), \textit{``Here everything's like um very flat. So I don't know where it is.''} (P12), and `\textit{`I didn't use this. Yeah, this part was well overwhelming''} (P7).
    
\end{itemize}

\subsection{Limitations and Future Work}
\label{subsec:limitations}

One of the issues that came up through the iterations is striking the right balance between the use case scenarios of \techname, specifically its use to explore a paper as an alternative to reading.
Similar ``distant reading'' approaches in the social sciences have received criticism for being suggested as objective alternatives to close reading, a practice considered integral to scholarship~\cite{Ascari2014dangers}.
In our studies, the expert review evaluation for the first iteration of \techname\ did not require participants to read the paper in advance.
Thus they spent more time using the system to understand the paper content---which was not the main focus of the system---than to generate and evaluate the summary.
The study setup following iteration 2 ensured that participants were already familiar with the paper, which allowed them to focus on the summary evaluation and customization tasks.
Participant reflections we saw in Sec.~\ref{subsec:summary_quality} and Sec.~\ref{subsec:insights} show that participants still used \techname\ as a way to check if they missed any important concepts, especially if they were unfamiliar with the domain of the paper.
Participant P3 suggested using \techname\ as a way to skim through papers so that ``\textit{if frequent concepts are not what I care, I can just leave this paper and turn to others.}''
On the other hand, comments about the disorienting effect of the paper layout in the paper info panel (see ``mental maps'' in Sec.~\ref{subsec:insights}) indicates that a better application of \techname\ would be toward supporting and summarization and  \textit{verification}, rather than exploration.
Integral to this approach would be to design a paper information view that preserves the appearance of the PDF view, thus preserving the reader's mental map and allowing them to build upon their close reading of the paper.

\edit{
The two test papers we chose for the user study were corresponding to the two different conditions---highly interdisciplinary papers spanning at least five domains and relatively typical CHI papers describing the algorithm, user study, and visualization design. 
Because of the authors’ limited knowledge background, we chose two CHI papers in which we had a better understanding and control of the content for our user study. 
We will eliminate this limitation by testing ConceptEVA on more diverse papers in the future. 
Besides, we are aware of the different summarization complexity for papers from different domains~\cite{Wu2021recursively, jain2021summarization, rauf2022meta, yang-etal-2018-aspect}, but consider it more of an NLP research problem rather than our main focus.
}

Participants also made suggestions for additional functions and features.
The most popular suggestions fell under the category of richer view coordination between the panels.
Specifically, participants suggested being able to support concept provenance and filtering within a selected section, or a direct linking between the summary text and the paper information panel.
However, this would also mean that \techname\ becomes more of an exploration tool providing an alternative to reading the paper rather than a support to summarize a paper, which is a different scope of work altogether, and a requirement that needs closer examination in terms of benefits and pitfalls.
On the other hand, other suggestions such as the one by P1 about being able to group concepts into groups relevant to the summary such as ``definition'', ``pipeline'', and ``preprocessing method''.
While the groups listed by P1 might work for a data science or data visualization domain, other domains might require entirely different groups than can then be examined to summarize contributions, offer critique, and present other salient ideas.
Allowing the user to create custom groups aided by additional NLP approaches like sentiment analysis and topic modeling could help users reflect on and critique the paper, and can be a helpful function to consider in a future iteration of the work.

Finally, a limitation of our study include technical issues such as network delays, rendering performance issues, and back-end computations to update concept embedding, sentence paraphrasing, or summary generation itself.
These, when they occurred, resulted in latency that influenced participants' experience and potentially their responses to questions like the NASA TLX scale.
While the focus of this work is not engineering or optimisation of the system, our future iterations will attempt to cut down performance or networking issues relating to latency.

%% file: sections/09-conclusion.tex
We have presented \techname, an interactive document summarization system aimed at long, and multi-domain documents of the kind seen in academic publications.
We show the iterative development and evaluation of \techname\ through two iterations.
The first iteration incorporates a hierarchical summarization technique with an interactive visualization of concepts extracted from the document using a reference ontology.
The second iteration, developed after evaluating the first iteration through an expert review, incorporates a multi-task longformer encoder decoder pre-trained for scientific documents that we fine-tune for paraphrasing and sentence embedding to handle long documents, and concepts visualized using a force-directed network that preserves semantic as well as co-occurrence relationships of document concepts.
We also introduce a ``focus-on'' function that allows users to choose concepts of interest, examine their relationship with co-occurring concepts, and choose relevant concepts that will then be incorporated into a custom summary.
An evaluation of \techname's second iteration through a within-subjects study using manually-generated summaries as baseline shows that \techname\ was helpful to participants for content-specific aspects of summarization, but participants with less experience struggled with critique-related aspects of summarization.
Participants largely preferred the summary created through \techname's human-in-the-loop approach over their own manually-generated summaries.
We discuss the implications of our findings and suggest future development and evaluations of mixed-initiative summarization systems.


%% file: main.bbl

\begin{thebibliography}{65}


\ifx \showCODEN    \undefined \def \showCODEN     #1{\unskip}     \fi
\ifx \showDOI      \undefined \def \showDOI       #1{#1}\fi
\ifx \showISBNx    \undefined \def \showISBNx     #1{\unskip}     \fi
\ifx \showISBNxiii \undefined \def \showISBNxiii  #1{\unskip}     \fi
\ifx \showISSN     \undefined \def \showISSN      #1{\unskip}     \fi
\ifx \showLCCN     \undefined \def \showLCCN      #1{\unskip}     \fi
\ifx \shownote     \undefined \def \shownote      #1{#1}          \fi
\ifx \showarticletitle \undefined \def \showarticletitle #1{#1}   \fi
\ifx \showURL      \undefined \def \showURL       {\relax}        \fi
\providecommand\bibfield[2]{#2}
\providecommand\bibinfo[2]{#2}
\providecommand\natexlab[1]{#1}
\providecommand\showeprint[2][]{arXiv:#2}

\bibitem[\protect\citeauthoryear{Ascari}{Ascari}{2014}]%
        {Ascari2014dangers}
\bibfield{author}{\bibinfo{person}{Maurizio Ascari}.}
  \bibinfo{year}{2014}\natexlab{}.
\newblock \showarticletitle{The Dangers of Distant Reading: Reassessing
  Moretti's Approach to Literary Genres}.
\newblock \bibinfo{journal}{\emph{Genre: Forms of Discourse and Culture}}
  \bibinfo{volume}{47}, \bibinfo{number}{1} (\bibinfo{year}{2014}),
  \bibinfo{pages}{1--19}.
\newblock
\urldef\tempurl%
\url{https://doi.org/10.1215/00166928-2392348}
\showDOI{\tempurl}


\bibitem[\protect\citeauthoryear{Auer, Bizer, Kobilarov, Lehmann, Cyganiak, and
  Ives}{Auer et~al\mbox{.}}{2007}]%
        {auer2007dbpedia}
\bibfield{author}{\bibinfo{person}{S{\"o}ren Auer}, \bibinfo{person}{Christian
  Bizer}, \bibinfo{person}{Georgi Kobilarov}, \bibinfo{person}{Jens Lehmann},
  \bibinfo{person}{Richard Cyganiak}, {and} \bibinfo{person}{Zachary Ives}.}
  \bibinfo{year}{2007}\natexlab{}.
\newblock \showarticletitle{Dbpedia: A nucleus for a web of open data}.
\newblock In \bibinfo{booktitle}{\emph{The semantic web}}.
  \bibinfo{publisher}{Springer}, \bibinfo{address}{Berlin, Heidelberg},
  \bibinfo{pages}{722--735}.
\newblock


\bibitem[\protect\citeauthoryear{Beltagy, Peters, and Cohan}{Beltagy
  et~al\mbox{.}}{2020}]%
        {Beltagy2020Longformer}
\bibfield{author}{\bibinfo{person}{Iz Beltagy}, \bibinfo{person}{Matthew~E.
  Peters}, {and} \bibinfo{person}{Arman Cohan}.}
  \bibinfo{year}{2020}\natexlab{}.
\newblock \showarticletitle{Longformer: The Long-Document Transformer}.
\newblock \bibinfo{journal}{\emph{CoRR}}  \bibinfo{volume}{abs/2004.05150}
  (\bibinfo{year}{2020}), 17.
\newblock
\urldef\tempurl%
\url{https://arxiv.org/abs/2004.05150}
\showURL{%
\tempurl}


\bibitem[\protect\citeauthoryear{Berger, McDonough, and Seversky}{Berger
  et~al\mbox{.}}{2016}]%
        {Berger2016cite2vec}
\bibfield{author}{\bibinfo{person}{Matthew Berger}, \bibinfo{person}{Katherine
  McDonough}, {and} \bibinfo{person}{Lee~M Seversky}.}
  \bibinfo{year}{2016}\natexlab{}.
\newblock \showarticletitle{cite2vec: Citation-driven document exploration via
  word embeddings}.
\newblock \bibinfo{journal}{\emph{IEEE transactions on visualization and
  computer graphics}} \bibinfo{volume}{23}, \bibinfo{number}{1}
  (\bibinfo{year}{2016}), \bibinfo{pages}{691--700}.
\newblock
\urldef\tempurl%
\url{https://doi.org/10.1109/TVCG.2016.2598667}
\showDOI{\tempurl}


\bibitem[\protect\citeauthoryear{Bollacker, Evans, Paritosh, Sturge, and
  Taylor}{Bollacker et~al\mbox{.}}{2008}]%
        {bollacker2008freebase}
\bibfield{author}{\bibinfo{person}{Kurt Bollacker}, \bibinfo{person}{Colin
  Evans}, \bibinfo{person}{Praveen Paritosh}, \bibinfo{person}{Tim Sturge},
  {and} \bibinfo{person}{Jamie Taylor}.} \bibinfo{year}{2008}\natexlab{}.
\newblock \showarticletitle{Freebase: a collaboratively created graph database
  for structuring human knowledge}. In \bibinfo{booktitle}{\emph{Proceedings of
  the ACM SIGMOD international conference on Management of data}}.
  \bibinfo{publisher}{ACM}, \bibinfo{address}{New York, NY},
  \bibinfo{pages}{1247--1250}.
\newblock
\urldef\tempurl%
\url{https://doi.org/10.1145/1376616.1376746}
\showDOI{\tempurl}


\bibitem[\protect\citeauthoryear{Bommasani and Cardie}{Bommasani and
  Cardie}{2020}]%
        {bommasani-cardie-2020-intrinsic}
\bibfield{author}{\bibinfo{person}{Rishi Bommasani} {and}
  \bibinfo{person}{Claire Cardie}.} \bibinfo{year}{2020}\natexlab{}.
\newblock \showarticletitle{Intrinsic Evaluation of Summarization Datasets}. In
  \bibinfo{booktitle}{\emph{Proceedings of the ACL Conference on Empirical
  Methods in Natural Language Processing}}. \bibinfo{publisher}{Association for
  Computational Linguistics}, \bibinfo{address}{Online},
  \bibinfo{pages}{8075--8096}.
\newblock
\urldef\tempurl%
\url{https://doi.org/10.18653/v1/2020.emnlp-main.649}
\showDOI{\tempurl}


\bibitem[\protect\citeauthoryear{Borgeaud, Mensch, Hoffmann, Cai, Rutherford,
  Millican, Van Den~Driessche, Lespiau, Damoc, Clark, De~Las~Casas, Guy,
  Menick, Ring, Hennigan, Huang, Maggiore, Jones, Cassirer, Brock, Paganini,
  Irving, Vinyals, Osindero, Simonyan, Rae, Elsen, and Sifre}{Borgeaud
  et~al\mbox{.}}{2022}]%
        {pmlr-v162-borgeaud22a}
\bibfield{author}{\bibinfo{person}{Sebastian Borgeaud}, \bibinfo{person}{Arthur
  Mensch}, \bibinfo{person}{Jordan Hoffmann}, \bibinfo{person}{Trevor Cai},
  \bibinfo{person}{Eliza Rutherford}, \bibinfo{person}{Katie Millican},
  \bibinfo{person}{George~Bm Van Den~Driessche}, \bibinfo{person}{Jean-Baptiste
  Lespiau}, \bibinfo{person}{Bogdan Damoc}, \bibinfo{person}{Aidan Clark},
  \bibinfo{person}{Diego De~Las~Casas}, \bibinfo{person}{Aurelia Guy},
  \bibinfo{person}{Jacob Menick}, \bibinfo{person}{Roman Ring},
  \bibinfo{person}{Tom Hennigan}, \bibinfo{person}{Saffron Huang},
  \bibinfo{person}{Loren Maggiore}, \bibinfo{person}{Chris Jones},
  \bibinfo{person}{Albin Cassirer}, \bibinfo{person}{Andy Brock},
  \bibinfo{person}{Michela Paganini}, \bibinfo{person}{Geoffrey Irving},
  \bibinfo{person}{Oriol Vinyals}, \bibinfo{person}{Simon Osindero},
  \bibinfo{person}{Karen Simonyan}, \bibinfo{person}{Jack Rae},
  \bibinfo{person}{Erich Elsen}, {and} \bibinfo{person}{Laurent Sifre}.}
  \bibinfo{year}{2022}\natexlab{}.
\newblock \showarticletitle{Improving Language Models by Retrieving from
  Trillions of Tokens}. In \bibinfo{booktitle}{\emph{Proceedings of the
  International Conference on Machine Learning}}
  \emph{(\bibinfo{series}{Proceedings of Machine Learning Research},
  Vol.~\bibinfo{volume}{162})}, \bibfield{editor}{\bibinfo{person}{Kamalika
  Chaudhuri}, \bibinfo{person}{Stefanie Jegelka}, \bibinfo{person}{Le~Song},
  \bibinfo{person}{Csaba Szepesvari}, \bibinfo{person}{Gang Niu}, {and}
  \bibinfo{person}{Sivan Sabato}} (Eds.). \bibinfo{publisher}{PMLR},
  \bibinfo{address}{Baltimore, Maryland}, \bibinfo{pages}{2206--2240}.
\newblock


\bibitem[\protect\citeauthoryear{Bowman, Angeli, Potts, and Manning}{Bowman
  et~al\mbox{.}}{2015}]%
        {snli:emnlp2015}
\bibfield{author}{\bibinfo{person}{Samuel~R. Bowman}, \bibinfo{person}{Gabor
  Angeli}, \bibinfo{person}{Christopher Potts}, {and}
  \bibinfo{person}{Christopher~D. Manning}.} \bibinfo{year}{2015}\natexlab{}.
\newblock \showarticletitle{A large annotated corpus for learning natural
  language inference}. In \bibinfo{booktitle}{\emph{Proceedings of the 2015
  Conference on Empirical Methods in Natural Language Processing}}.
  \bibinfo{publisher}{Association for Computational Linguistics},
  \bibinfo{address}{Lisbon, Portugal}, \bibinfo{pages}{632--642}.
\newblock


\bibitem[\protect\citeauthoryear{Chandrasegaran, Bryan, Shidara, Chuan, and
  Ma}{Chandrasegaran et~al\mbox{.}}{2019}]%
        {Chandrasegaran2019talktraces}
\bibfield{author}{\bibinfo{person}{Senthil Chandrasegaran},
  \bibinfo{person}{Chris Bryan}, \bibinfo{person}{Hidekazu Shidara},
  \bibinfo{person}{Tun-Yeng Chuan}, {and} \bibinfo{person}{Kwan-Liu Ma}.}
  \bibinfo{year}{2019}\natexlab{}.
\newblock \showarticletitle{{TalkTraces}: Real-Time Capture and Visualization
  of Verbal Content in Meetings}. In \bibinfo{booktitle}{\emph{Proceedings of
  the ACM CHI Conference on Human Factors in Computing Systems}}.
  \bibinfo{publisher}{ACM}, \bibinfo{address}{New York, NY},
  \bibinfo{pages}{577:1--577:14}.
\newblock
\urldef\tempurl%
\url{https://doi.org/10.1145/3290605.3300807}
\showDOI{\tempurl}


\bibitem[\protect\citeauthoryear{Clement, Bierbaum, O'Keeffe, and
  Alemi}{Clement et~al\mbox{.}}{2019}]%
        {clement2019arxiv}
\bibfield{author}{\bibinfo{person}{Colin~B. Clement}, \bibinfo{person}{Matthew
  Bierbaum}, \bibinfo{person}{Kevin~P. O'Keeffe}, {and}
  \bibinfo{person}{Alexander~A. Alemi}.} \bibinfo{year}{2019}\natexlab{}.
\newblock \showarticletitle{On the Use of ArXiv as a Dataset}.
\newblock \bibinfo{journal}{\emph{CoRR}}  \bibinfo{volume}{abs/1905.00075}
  (\bibinfo{year}{2019}), 7.
\newblock
\urldef\tempurl%
\url{https://arxiv.org/abs/1905.00075}
\showURL{%
\tempurl}


\bibitem[\protect\citeauthoryear{Cohan, Feldman, Beltagy, Downey, and
  Weld}{Cohan et~al\mbox{.}}{2020}]%
        {Cohan2020specter}
\bibfield{author}{\bibinfo{person}{Arman Cohan}, \bibinfo{person}{Sergey
  Feldman}, \bibinfo{person}{Iz Beltagy}, \bibinfo{person}{Doug Downey}, {and}
  \bibinfo{person}{Daniel~S Weld}.} \bibinfo{year}{2020}\natexlab{}.
\newblock \showarticletitle{{SPECTER:} Document-level Representation Learning
  using Citation-informed Transformers}. In
  \bibinfo{booktitle}{\emph{Proceedings of the 58th Annual Meeting of the
  Association for Computational Linguistics}}. \bibinfo{publisher}{Association
  for Computational Linguistics}, \bibinfo{address}{Online},
  \bibinfo{pages}{2270--2282}.
\newblock
\urldef\tempurl%
\url{https://doi.org/10.18653/v1/2020.acl-main.207}
\showDOI{\tempurl}


\bibitem[\protect\citeauthoryear{Devlin, Chang, Lee, and Toutanova}{Devlin
  et~al\mbox{.}}{2019}]%
        {devlin2019bert}
\bibfield{author}{\bibinfo{person}{Jacob Devlin}, \bibinfo{person}{Ming-Wei
  Chang}, \bibinfo{person}{Kenton Lee}, {and} \bibinfo{person}{Kristina
  Toutanova}.} \bibinfo{year}{2019}\natexlab{}.
\newblock \bibinfo{title}{{BERT}: Pre-training of Deep Bidirectional
  Transformers for Language Understanding}.
\newblock , \bibinfo{numpages}{4171--4186}~pages.
\newblock
\urldef\tempurl%
\url{https://doi.org/10.18653/v1/N19-1423}
\showDOI{\tempurl}


\bibitem[\protect\citeauthoryear{Dorr, Monz, President, Schwartz, and
  Zajic}{Dorr et~al\mbox{.}}{2005}]%
        {Dorr2005methodology}
\bibfield{author}{\bibinfo{person}{Bonnie Dorr}, \bibinfo{person}{Christof
  Monz}, \bibinfo{person}{Stacy President}, \bibinfo{person}{Richard Schwartz},
  {and} \bibinfo{person}{David Zajic}.} \bibinfo{year}{2005}\natexlab{}.
\newblock \showarticletitle{A methodology for extrinsic evaluation of text
  summarization: does ROUGE correlate?}. In
  \bibinfo{booktitle}{\emph{Proceedings of the ACL Workshop on Intrinsic and
  Extrinsic Evaluation Measures for Machine Translation and/or Summarization}}.
  \bibinfo{publisher}{Association for Computational Linguistics},
  \bibinfo{address}{Ann Arbor, Michigan}, \bibinfo{pages}{1--8}.
\newblock


\bibitem[\protect\citeauthoryear{El-Assady, Gold, Acevedo, Collins, and
  Keim}{El-Assady et~al\mbox{.}}{2016}]%
        {El2016contovi}
\bibfield{author}{\bibinfo{person}{Mennatallah El-Assady},
  \bibinfo{person}{Valentin Gold}, \bibinfo{person}{Carmela Acevedo},
  \bibinfo{person}{Christopher Collins}, {and} \bibinfo{person}{Daniel Keim}.}
  \bibinfo{year}{2016}\natexlab{}.
\newblock \showarticletitle{ConToVi: Multi-party conversation exploration using
  topic-space views}.
\newblock \bibinfo{journal}{\emph{Computer Graphics Forum}}
  \bibinfo{volume}{35}, \bibinfo{number}{3} (\bibinfo{year}{2016}),
  \bibinfo{pages}{431--440}.
\newblock
\urldef\tempurl%
\url{https://doi.org/10.1111/cgf.12919}
\showDOI{\tempurl}


\bibitem[\protect\citeauthoryear{El-Assady, Sevastjanova, Gipp, Keim, and
  Collins}{El-Assady et~al\mbox{.}}{2017}]%
        {El2017nerex}
\bibfield{author}{\bibinfo{person}{Mennatallah El-Assady},
  \bibinfo{person}{Rita Sevastjanova}, \bibinfo{person}{Bela Gipp},
  \bibinfo{person}{Daniel Keim}, {and} \bibinfo{person}{Christopher Collins}.}
  \bibinfo{year}{2017}\natexlab{}.
\newblock \showarticletitle{NEREx: Named-Entity Relationship Exploration in
  Multi-Party Conversations}.
\newblock \bibinfo{journal}{\emph{Computer Graphics Forum}}
  \bibinfo{volume}{36}, \bibinfo{number}{3} (\bibinfo{year}{2017}),
  \bibinfo{pages}{213--225}.
\newblock
\urldef\tempurl%
\url{https://doi.org/10.1111/cgf.13181}
\showDOI{\tempurl}


\bibitem[\protect\citeauthoryear{El-Kassas, Salama, Rafea, and
  Mohamed}{El-Kassas et~al\mbox{.}}{2021}]%
        {El2021automatic}
\bibfield{author}{\bibinfo{person}{Wafaa~S El-Kassas},
  \bibinfo{person}{Cherif~R Salama}, \bibinfo{person}{Ahmed~A Rafea}, {and}
  \bibinfo{person}{Hoda~K Mohamed}.} \bibinfo{year}{2021}\natexlab{}.
\newblock \showarticletitle{Automatic text summarization: A comprehensive
  survey}.
\newblock \bibinfo{journal}{\emph{Expert Systems with Applications}}
  \bibinfo{volume}{165} (\bibinfo{year}{2021}), \bibinfo{pages}{113679}.
\newblock
\urldef\tempurl%
\url{https://doi.org/10.1016/j.eswa.2020.113679}
\showDOI{\tempurl}


\bibitem[\protect\citeauthoryear{Gupta and Gupta}{Gupta and Gupta}{2019}]%
        {Gupta2019abstractive}
\bibfield{author}{\bibinfo{person}{Som Gupta} {and}
  \bibinfo{person}{Sanjai~Kumar Gupta}.} \bibinfo{year}{2019}\natexlab{}.
\newblock \showarticletitle{Abstractive summarization: An overview of the state
  of the art}.
\newblock \bibinfo{journal}{\emph{Expert Systems with Applications}}
  \bibinfo{volume}{121} (\bibinfo{year}{2019}), \bibinfo{pages}{49--65}.
\newblock
\urldef\tempurl%
\url{https://doi.org/10.1016/j.eswa.2018.12.011}
\showDOI{\tempurl}


\bibitem[\protect\citeauthoryear{Gupta and Lehal}{Gupta and Lehal}{2010}]%
        {Gupta2010survey}
\bibfield{author}{\bibinfo{person}{Vishal Gupta} {and}
  \bibinfo{person}{Gurpreet~Singh Lehal}.} \bibinfo{year}{2010}\natexlab{}.
\newblock \showarticletitle{A survey of text summarization extractive
  techniques}.
\newblock \bibinfo{journal}{\emph{Journal of emerging technologies in web
  intelligence}} \bibinfo{volume}{2}, \bibinfo{number}{3}
  (\bibinfo{year}{2010}), \bibinfo{pages}{258--268}.
\newblock
\urldef\tempurl%
\url{https://doi.org/10.4304/jetwi.2.3.258-268}
\showDOI{\tempurl}


\bibitem[\protect\citeauthoryear{Hariharan and Srinivasan}{Hariharan and
  Srinivasan}{2010}]%
        {hariharan2010studies}
\bibfield{author}{\bibinfo{person}{Shanmugasundaram Hariharan} {and}
  \bibinfo{person}{Rengaramanujam Srinivasan}.}
  \bibinfo{year}{2010}\natexlab{}.
\newblock \showarticletitle{Studies on intrinsic summary evaluation}.
\newblock \bibinfo{journal}{\emph{International Journal of Artificial
  Intelligence and Soft Computing}} \bibinfo{volume}{2}, \bibinfo{number}{1-2}
  (\bibinfo{year}{2010}), \bibinfo{pages}{58--76}.
\newblock
\urldef\tempurl%
\url{https://doi.org/10.1504/IJAISC.2010.032513}
\showDOI{\tempurl}


\bibitem[\protect\citeauthoryear{Hart and Staveland}{Hart and
  Staveland}{1988}]%
        {Hart1988development}
\bibfield{author}{\bibinfo{person}{Sandra~G Hart} {and}
  \bibinfo{person}{Lowell~E Staveland}.} \bibinfo{year}{1988}\natexlab{}.
\newblock \showarticletitle{Development of NASA-TLX (Task Load Index): Results
  of empirical and theoretical research}.
\newblock In \bibinfo{booktitle}{\emph{Human Mental Workload}}.
  \bibinfo{series}{Advances in psychology}, Vol.~\bibinfo{volume}{52}.
  \bibinfo{publisher}{Elsevier}, \bibinfo{address}{Amsterdam, Netherlands},
  \bibinfo{pages}{139--183}.
\newblock
\urldef\tempurl%
\url{https://doi.org/10.1016/S0166-4115(08)62386-9}
\showDOI{\tempurl}


\bibitem[\protect\citeauthoryear{Heimerl and Gleicher}{Heimerl and
  Gleicher}{2018}]%
        {heimerl2018interactive}
\bibfield{author}{\bibinfo{person}{Florian Heimerl} {and}
  \bibinfo{person}{Michael Gleicher}.} \bibinfo{year}{2018}\natexlab{}.
\newblock \showarticletitle{Interactive analysis of word vector embeddings}.
\newblock \bibinfo{journal}{\emph{Computer Graphics Forum}}
  \bibinfo{volume}{37}, \bibinfo{number}{3} (\bibinfo{year}{2018}),
  \bibinfo{pages}{253--265}.
\newblock
\urldef\tempurl%
\url{https://doi.org/10.1111/cgf.13417}
\showDOI{\tempurl}


\bibitem[\protect\citeauthoryear{Hirao, Sasaki, and Isozaki}{Hirao
  et~al\mbox{.}}{2001}]%
        {Hirao2001AnEE}
\bibfield{author}{\bibinfo{person}{Tsutomu Hirao}, \bibinfo{person}{Yutaka
  Sasaki}, {and} \bibinfo{person}{Hideki Isozaki}.}
  \bibinfo{year}{2001}\natexlab{}.
\newblock \showarticletitle{An extrinsic evaluation for question-biased text
  summarization on {QA} tasks}. In \bibinfo{booktitle}{\emph{Proceedings of the
  NAACL Workshop on Automatic Summarization}}. \bibinfo{publisher}{Association
  for Computational Linguistics}, \bibinfo{address}{Pittsburgh, PA},
  \bibinfo{pages}{61--68}.
\newblock


\bibitem[\protect\citeauthoryear{Jain, Borah, and Biswas}{Jain
  et~al\mbox{.}}{2021}]%
        {jain2021summarization}
\bibfield{author}{\bibinfo{person}{Deepali Jain}, \bibinfo{person}{Malaya~Dutta
  Borah}, {and} \bibinfo{person}{Anupam Biswas}.}
  \bibinfo{year}{2021}\natexlab{}.
\newblock \showarticletitle{Summarization of legal documents: Where are we now
  and the way forward}.
\newblock \bibinfo{journal}{\emph{Computer Science Review}}
  \bibinfo{volume}{40} (\bibinfo{year}{2021}), \bibinfo{pages}{100388}.
\newblock
\urldef\tempurl%
\url{https://doi.org/10.1016/j.cosrev.2021.100388}
\showDOI{\tempurl}


\bibitem[\protect\citeauthoryear{Johnson, Douze, and J{\'e}gou}{Johnson
  et~al\mbox{.}}{2019}]%
        {johnson2019billion}
\bibfield{author}{\bibinfo{person}{Jeff Johnson}, \bibinfo{person}{Matthijs
  Douze}, {and} \bibinfo{person}{Herv{\'e} J{\'e}gou}.}
  \bibinfo{year}{2019}\natexlab{}.
\newblock \showarticletitle{Billion-scale similarity search with {GPUs}}.
\newblock \bibinfo{journal}{\emph{IEEE Transactions on Big Data}}
  \bibinfo{volume}{7}, \bibinfo{number}{3} (\bibinfo{year}{2019}),
  \bibinfo{pages}{535--547}.
\newblock
\urldef\tempurl%
\url{https://doi.org/10.1109/TBDATA.2019.2921572}
\showDOI{\tempurl}


\bibitem[\protect\citeauthoryear{Khandelwal, Levy, Jurafsky, Zettlemoyer, and
  Lewis}{Khandelwal et~al\mbox{.}}{2020}]%
        {Khandelwal2020Generalization}
\bibfield{author}{\bibinfo{person}{Urvashi Khandelwal}, \bibinfo{person}{Omer
  Levy}, \bibinfo{person}{Dan Jurafsky}, \bibinfo{person}{Luke Zettlemoyer},
  {and} \bibinfo{person}{Mike Lewis}.} \bibinfo{year}{2020}\natexlab{}.
\newblock \showarticletitle{Generalization through Memorization: Nearest
  Neighbor Language Models}. In \bibinfo{booktitle}{\emph{International
  Conference on Learning Representations}}.
  \bibinfo{publisher}{OpenReview.net}, \bibinfo{address}{Addis Ababa,
  Ethiopia}, 13.
\newblock


\bibitem[\protect\citeauthoryear{Kim, Hoque, Kim, and Agrawala}{Kim
  et~al\mbox{.}}{2018}]%
        {kim2018facilitating}
\bibfield{author}{\bibinfo{person}{Dae~Hyun Kim}, \bibinfo{person}{Enamul
  Hoque}, \bibinfo{person}{Juho Kim}, {and} \bibinfo{person}{Maneesh
  Agrawala}.} \bibinfo{year}{2018}\natexlab{}.
\newblock \showarticletitle{Facilitating document reading by linking text and
  tables}. In \bibinfo{booktitle}{\emph{Proceedings of the ACM Symposium on
  User Interface Software and Technology}}. \bibinfo{publisher}{Association for
  Computing Machinery}, \bibinfo{address}{New York, NY},
  \bibinfo{pages}{423--434}.
\newblock
\urldef\tempurl%
\url{https://doi.org/10.1145/3242587.3242617}
\showDOI{\tempurl}


\bibitem[\protect\citeauthoryear{Kucher, Paradis, and Kerren}{Kucher
  et~al\mbox{.}}{2018}]%
        {Kucher2018state}
\bibfield{author}{\bibinfo{person}{Kostiantyn Kucher}, \bibinfo{person}{Carita
  Paradis}, {and} \bibinfo{person}{Andreas Kerren}.}
  \bibinfo{year}{2018}\natexlab{}.
\newblock \showarticletitle{The state of the art in sentiment visualization}.
\newblock \bibinfo{journal}{\emph{Computer Graphics Forum}}
  \bibinfo{volume}{37}, \bibinfo{number}{1} (\bibinfo{year}{2018}),
  \bibinfo{pages}{71--96}.
\newblock
\urldef\tempurl%
\url{https://doi.org/10.1111/cgf.13217}
\showDOI{\tempurl}


\bibitem[\protect\citeauthoryear{Lai, Smith-Renner, Zhang, Cheng, Zhang,
  Tetreault, and Jaimes-Larrarte}{Lai et~al\mbox{.}}{2022}]%
        {lai2022exploration}
\bibfield{author}{\bibinfo{person}{Vivian Lai}, \bibinfo{person}{Alison
  Smith-Renner}, \bibinfo{person}{Ke Zhang}, \bibinfo{person}{Ruijia Cheng},
  \bibinfo{person}{Wenjuan Zhang}, \bibinfo{person}{Joel Tetreault}, {and}
  \bibinfo{person}{Alejandro Jaimes-Larrarte}.}
  \bibinfo{year}{2022}\natexlab{}.
\newblock \showarticletitle{An Exploration of Post-Editing Effectiveness in
  Text Summarization}. In \bibinfo{booktitle}{\emph{Proceedings of the
  Conference of the North American Chapter of the Association for Computational
  Linguistics: Human Language Technologies}}. \bibinfo{publisher}{Association
  for Computational Linguistics}, \bibinfo{address}{Seattle, United States},
  \bibinfo{pages}{475--493}.
\newblock
\urldef\tempurl%
\url{https://doi.org/10.18653/v1/2022.naacl-main.35}
\showDOI{\tempurl}


\bibitem[\protect\citeauthoryear{Lewis, Liu, Goyal, Ghazvininejad, Mohamed,
  Levy, Stoyanov, and Zettlemoyer}{Lewis et~al\mbox{.}}{2020}]%
        {lewis2019bart}
\bibfield{author}{\bibinfo{person}{Mike Lewis}, \bibinfo{person}{Yinhan Liu},
  \bibinfo{person}{Naman Goyal}, \bibinfo{person}{Marjan Ghazvininejad},
  \bibinfo{person}{Abdelrahman Mohamed}, \bibinfo{person}{Omer Levy},
  \bibinfo{person}{Veselin Stoyanov}, {and} \bibinfo{person}{Luke
  Zettlemoyer}.} \bibinfo{year}{2020}\natexlab{}.
\newblock \showarticletitle{{BART}: Denoising Sequence-to-Sequence Pre-training
  for Natural Language Generation, Translation, and Comprehension}. In
  \bibinfo{booktitle}{\emph{Proceedings of the Annual Meeting of the
  Association for Computational Linguistics}}. \bibinfo{publisher}{Association
  for Computational Linguistics}, \bibinfo{address}{Online},
  \bibinfo{pages}{7871--7880}.
\newblock
\urldef\tempurl%
\url{https://doi.org/10.18653/v1/2020.acl-main.703}
\showDOI{\tempurl}


\bibitem[\protect\citeauthoryear{Li, Zhu, Zhang, Zong, and He}{Li
  et~al\mbox{.}}{2020}]%
        {Li_Zhu_Zhang_Zong_He_2020}
\bibfield{author}{\bibinfo{person}{Haoran Li}, \bibinfo{person}{Junnan Zhu},
  \bibinfo{person}{Jiajun Zhang}, \bibinfo{person}{Chengqing Zong}, {and}
  \bibinfo{person}{Xiaodong He}.} \bibinfo{year}{2020}\natexlab{}.
\newblock \showarticletitle{Keywords-Guided Abstractive Sentence
  Summarization}.
\newblock \bibinfo{journal}{\emph{Proceedings of the AAAI Conference on
  Artificial Intelligence}} \bibinfo{volume}{34}, \bibinfo{number}{05}
  (\bibinfo{year}{2020}), \bibinfo{pages}{8196--8203}.
\newblock
\urldef\tempurl%
\url{https://doi.org/10.1609/aaai.v34i05.6333}
\showDOI{\tempurl}


\bibitem[\protect\citeauthoryear{Lin, Ford, Adar, and Hecht}{Lin
  et~al\mbox{.}}{2018}]%
        {lin2018vizbywiki}
\bibfield{author}{\bibinfo{person}{Allen~Yilun Lin}, \bibinfo{person}{Joshua
  Ford}, \bibinfo{person}{Eytan Adar}, {and} \bibinfo{person}{Brent Hecht}.}
  \bibinfo{year}{2018}\natexlab{}.
\newblock \showarticletitle{{VizByWiki:} Mining data visualizations from the
  web to enrich news articles}. In \bibinfo{booktitle}{\emph{Proceedings of the
  World Wide Web Conference}}. \bibinfo{publisher}{ACM}, \bibinfo{address}{New
  York, NY}, \bibinfo{pages}{873--882}.
\newblock
\urldef\tempurl%
\url{https://doi.org/10.1145/3178876.3186135}
\showDOI{\tempurl}


\bibitem[\protect\citeauthoryear{Lin}{Lin}{2004}]%
        {lin-2004-rouge}
\bibfield{author}{\bibinfo{person}{Chin-Yew Lin}.}
  \bibinfo{year}{2004}\natexlab{}.
\newblock \showarticletitle{{ROUGE}: A Package for Automatic Evaluation of
  Summaries}. In \bibinfo{booktitle}{\emph{Text Summarization Branches Out}}.
  \bibinfo{publisher}{Association for Computational Linguistics},
  \bibinfo{address}{Barcelona, Spain}, \bibinfo{pages}{74--81}.
\newblock


\bibitem[\protect\citeauthoryear{Lin and Hovy}{Lin and Hovy}{2000}]%
        {Lin2000automated}
\bibfield{author}{\bibinfo{person}{Chin-Yew Lin} {and} \bibinfo{person}{Eduard
  Hovy}.} \bibinfo{year}{2000}\natexlab{}.
\newblock \showarticletitle{The automated acquisition of topic signatures for
  text summarization}. In \bibinfo{booktitle}{\emph{{COLING} 2000 Volume 1: The
  18th International Conference on Computational Linguistics}}.
  \bibinfo{publisher}{Association for Computational Linguistics},
  \bibinfo{address}{Online}, \bibinfo{pages}{495--501}.
\newblock
\urldef\tempurl%
\url{https://doi.org/10.3115/990820.990892}
\showDOI{\tempurl}


\bibitem[\protect\citeauthoryear{Liu, Bremer, Thiagarajan, Srikumar, Wang,
  Livnat, and Pascucci}{Liu et~al\mbox{.}}{2017}]%
        {liu2017visual}
\bibfield{author}{\bibinfo{person}{Shusen Liu}, \bibinfo{person}{Peer-Timo
  Bremer}, \bibinfo{person}{Jayaraman~J Thiagarajan}, \bibinfo{person}{Vivek
  Srikumar}, \bibinfo{person}{Bei Wang}, \bibinfo{person}{Yarden Livnat}, {and}
  \bibinfo{person}{Valerio Pascucci}.} \bibinfo{year}{2017}\natexlab{}.
\newblock \showarticletitle{Visual exploration of semantic relationships in
  neural word embeddings}.
\newblock \bibinfo{journal}{\emph{IEEE transactions on visualization and
  computer graphics}} \bibinfo{volume}{24}, \bibinfo{number}{1}
  (\bibinfo{year}{2017}), \bibinfo{pages}{553--562}.
\newblock
\urldef\tempurl%
\url{https://doi.org/10.1109/TVCG.2017.2745141}
\showDOI{\tempurl}


\bibitem[\protect\citeauthoryear{Luhn}{Luhn}{1958}]%
        {Luhn1958automatic}
\bibfield{author}{\bibinfo{person}{Hans~Peter Luhn}.}
  \bibinfo{year}{1958}\natexlab{}.
\newblock \showarticletitle{The automatic creation of literature abstracts}.
\newblock \bibinfo{journal}{\emph{IBM Journal of research and development}}
  \bibinfo{volume}{2}, \bibinfo{number}{2} (\bibinfo{year}{1958}),
  \bibinfo{pages}{159--165}.
\newblock
\urldef\tempurl%
\url{https://doi.org/10.1147/rd.22.0159}
\showDOI{\tempurl}


\bibitem[\protect\citeauthoryear{McInnes, Healy, and Melville}{McInnes
  et~al\mbox{.}}{2018}]%
        {mcinnes2020umap}
\bibfield{author}{\bibinfo{person}{Leland McInnes}, \bibinfo{person}{John
  Healy}, {and} \bibinfo{person}{James Melville}.}
  \bibinfo{year}{2018}\natexlab{}.
\newblock \showarticletitle{UMAP: Uniform Manifold Approximation and Projection
  for Dimension Reduction}.
\newblock \bibinfo{journal}{\emph{{CoRR}}}  \bibinfo{volume}{1802.03426}
  (\bibinfo{year}{2018}), 63.
\newblock
\urldef\tempurl%
\url{https://doi.org/10.48550/ARXIV.1802.03426}
\showDOI{\tempurl}


\bibitem[\protect\citeauthoryear{Mendes, Jakob, Garc\'{\i}a-Silva, and
  Bizer}{Mendes et~al\mbox{.}}{2011}]%
        {10.1145/2063518.2063519}
\bibfield{author}{\bibinfo{person}{Pablo~N. Mendes}, \bibinfo{person}{Max
  Jakob}, \bibinfo{person}{Andr\'{e}s Garc\'{\i}a-Silva}, {and}
  \bibinfo{person}{Christian Bizer}.} \bibinfo{year}{2011}\natexlab{}.
\newblock \showarticletitle{DBpedia Spotlight: Shedding Light on the Web of
  Documents}. In \bibinfo{booktitle}{\emph{Proceedings of the ACM International
  Conference on Semantic Systems}}. \bibinfo{publisher}{Association for
  Computing Machinery}, \bibinfo{address}{New York, NY, USA},
  \bibinfo{pages}{1–8}.
\newblock
\urldef\tempurl%
\url{https://doi.org/10.1145/2063518.2063519}
\showDOI{\tempurl}


\bibitem[\protect\citeauthoryear{Miller}{Miller}{2019}]%
        {bert-extractive-summarizer}
\bibfield{author}{\bibinfo{person}{Derek Miller}.}
  \bibinfo{year}{2019}\natexlab{}.
\newblock \bibinfo{title}{{BERT} Extractive Summarizer}.
\newblock
  \bibinfo{howpublished}{{https://github.com/dmmiller612/bert-extractive-summarizer}}.
\newblock


\bibitem[\protect\citeauthoryear{Moramarco, Papadopoulos~Korfiatis, Savkov, and
  Reiter}{Moramarco et~al\mbox{.}}{2021}]%
        {moramarco2021preliminary}
\bibfield{author}{\bibinfo{person}{Francesco Moramarco}, \bibinfo{person}{Alex
  Papadopoulos~Korfiatis}, \bibinfo{person}{Aleksandar Savkov}, {and}
  \bibinfo{person}{Ehud Reiter}.} \bibinfo{year}{2021}\natexlab{}.
\newblock \showarticletitle{A Preliminary Study on Evaluating Consultation
  Notes With Post-Editing}. In \bibinfo{booktitle}{\emph{Proceedings of the
  Workshop on Human Evaluation of NLP Systems (HumEval)}}.
  \bibinfo{publisher}{Association for Computational Linguistics},
  \bibinfo{address}{Online}, \bibinfo{pages}{62--68}.
\newblock


\bibitem[\protect\citeauthoryear{Munzner}{Munzner}{2009}]%
        {munzner2009nested}
\bibfield{author}{\bibinfo{person}{Tamara Munzner}.}
  \bibinfo{year}{2009}\natexlab{}.
\newblock \showarticletitle{A nested model for visualization design and
  validation}.
\newblock \bibinfo{journal}{\emph{IEEE transactions on visualization and
  computer graphics}} \bibinfo{volume}{15}, \bibinfo{number}{6}
  (\bibinfo{year}{2009}), \bibinfo{pages}{921--928}.
\newblock
\urldef\tempurl%
\url{https://doi.org/10.1109/TVCG.2009.111}
\showDOI{\tempurl}


\bibitem[\protect\citeauthoryear{Murray, Kleinbauer, Poller, Becker, Renals,
  and Kilgour}{Murray et~al\mbox{.}}{2009}]%
        {murray2009extrinsic}
\bibfield{author}{\bibinfo{person}{Gabriel Murray}, \bibinfo{person}{Thomas
  Kleinbauer}, \bibinfo{person}{Peter Poller}, \bibinfo{person}{Tilman Becker},
  \bibinfo{person}{Steve Renals}, {and} \bibinfo{person}{Jonathan Kilgour}.}
  \bibinfo{year}{2009}\natexlab{}.
\newblock \showarticletitle{Extrinsic summarization evaluation: A decision
  audit task}.
\newblock \bibinfo{journal}{\emph{ACM Transactions on Speech and Language
  Processing}} \bibinfo{volume}{6}, \bibinfo{number}{2} (\bibinfo{year}{2009}),
  \bibinfo{pages}{1--29}.
\newblock
\urldef\tempurl%
\url{https://doi.org/10.1145/1596517.1596518}
\showDOI{\tempurl}


\bibitem[\protect\citeauthoryear{Narechania, Karduni, Wesslen, and
  Wall}{Narechania et~al\mbox{.}}{2021}]%
        {Narechania2021vitality}
\bibfield{author}{\bibinfo{person}{Arpit Narechania}, \bibinfo{person}{Alireza
  Karduni}, \bibinfo{person}{Ryan Wesslen}, {and} \bibinfo{person}{Emily
  Wall}.} \bibinfo{year}{2021}\natexlab{}.
\newblock \showarticletitle{{vitaLITy:} Promoting Serendipitous Discovery of
  Academic Literature with Transformers \& Visual Analytics}.
\newblock \bibinfo{journal}{\emph{IEEE Transactions on Visualization and
  Computer Graphics}} \bibinfo{volume}{28}, \bibinfo{number}{1}
  (\bibinfo{year}{2021}), \bibinfo{pages}{486--496}.
\newblock
\urldef\tempurl%
\url{https://doi.org/10.1109/TVCG.2021.3114820}
\showDOI{\tempurl}


\bibitem[\protect\citeauthoryear{Norooz, Mauriello, Jorgensen, McNally, and
  Froehlich}{Norooz et~al\mbox{.}}{2015}]%
        {Norooz2015bodyvis}
\bibfield{author}{\bibinfo{person}{Leyla Norooz},
  \bibinfo{person}{Matthew~Louis Mauriello}, \bibinfo{person}{Anita Jorgensen},
  \bibinfo{person}{Brenna McNally}, {and} \bibinfo{person}{Jon~E Froehlich}.}
  \bibinfo{year}{2015}\natexlab{}.
\newblock \showarticletitle{{BodyVis:} A new approach to body learning through
  wearable sensing and visualization}. In \bibinfo{booktitle}{\emph{Proceedings
  of the ACM CHI Conference on Human Factors in Computing Systems}}.
  \bibinfo{publisher}{ACM}, \bibinfo{address}{New York, NY},
  \bibinfo{pages}{1025--1034}.
\newblock
\urldef\tempurl%
\url{https://doi.org/10.1145/2702123.2702299}
\showDOI{\tempurl}


\bibitem[\protect\citeauthoryear{Park, Das, Duggal, Wright, Shaikh, Hohman, and
  Chau}{Park et~al\mbox{.}}{2021}]%
        {Park2021neurocartography}
\bibfield{author}{\bibinfo{person}{Haekyu Park}, \bibinfo{person}{Nilaksh Das},
  \bibinfo{person}{Rahul Duggal}, \bibinfo{person}{Austin~P Wright},
  \bibinfo{person}{Omar Shaikh}, \bibinfo{person}{Fred Hohman}, {and}
  \bibinfo{person}{Duen Horng~Polo Chau}.} \bibinfo{year}{2021}\natexlab{}.
\newblock \showarticletitle{NeuroCartography: Scalable Automatic Visual
  Summarization of Concepts in Deep Neural Networks}.
\newblock \bibinfo{journal}{\emph{IEEE Transactions on Visualization and
  Computer Graphics}} \bibinfo{volume}{28}, \bibinfo{number}{1}
  (\bibinfo{year}{2021}), \bibinfo{pages}{813--823}.
\newblock
\urldef\tempurl%
\url{https://doi.org/10.1109/TVCG.2021.3114858}
\showDOI{\tempurl}


\bibitem[\protect\citeauthoryear{Radford, Wu, Child, Luan, Amodei, and
  Sutskever}{Radford et~al\mbox{.}}{2019}]%
        {radford2019language}
\bibfield{author}{\bibinfo{person}{Alec Radford}, \bibinfo{person}{Jeff Wu},
  \bibinfo{person}{Rewon Child}, \bibinfo{person}{David Luan},
  \bibinfo{person}{Dario Amodei}, {and} \bibinfo{person}{Ilya Sutskever}.}
  \bibinfo{year}{2019}\natexlab{}.
\newblock \bibinfo{title}{Language Models are Unsupervised Multitask Learners}.
\newblock
\newblock


\bibitem[\protect\citeauthoryear{Raffel, Shazeer, Roberts, Lee, Narang, Matena,
  Zhou, Li, and Liu}{Raffel et~al\mbox{.}}{2020}]%
        {raffel2020exploring}
\bibfield{author}{\bibinfo{person}{Colin Raffel}, \bibinfo{person}{Noam
  Shazeer}, \bibinfo{person}{Adam Roberts}, \bibinfo{person}{Katherine Lee},
  \bibinfo{person}{Sharan Narang}, \bibinfo{person}{Michael Matena},
  \bibinfo{person}{Yanqi Zhou}, \bibinfo{person}{Wei Li}, {and}
  \bibinfo{person}{Peter~J Liu}.} \bibinfo{year}{2020}\natexlab{}.
\newblock \showarticletitle{Exploring the limits of transfer learning with a
  unified text-to-text transformer}.
\newblock \bibinfo{journal}{\emph{The Journal of Machine Learning Research}}
  \bibinfo{volume}{21}, \bibinfo{number}{1} (\bibinfo{year}{2020}),
  \bibinfo{pages}{5485--5551}.
\newblock
\urldef\tempurl%
\url{https://doi.org/10.5555/3455716.3455856}
\showDOI{\tempurl}


\bibitem[\protect\citeauthoryear{Rahm and Bernstein}{Rahm and
  Bernstein}{2001}]%
        {rahm2001survey}
\bibfield{author}{\bibinfo{person}{Erhard Rahm} {and} \bibinfo{person}{Philip~A
  Bernstein}.} \bibinfo{year}{2001}\natexlab{}.
\newblock \showarticletitle{A survey of approaches to automatic schema
  matching}.
\newblock \bibinfo{journal}{\emph{the VLDB Journal}} \bibinfo{volume}{10},
  \bibinfo{number}{4} (\bibinfo{year}{2001}), \bibinfo{pages}{334--350}.
\newblock
\urldef\tempurl%
\url{https://doi.org/10.1007/S007780100057}
\showDOI{\tempurl}


\bibitem[\protect\citeauthoryear{Rauf, Pad{\'o}, and Pradel}{Rauf
  et~al\mbox{.}}{2022}]%
        {rauf2022meta}
\bibfield{author}{\bibinfo{person}{Moiz Rauf}, \bibinfo{person}{Sebastian
  Pad{\'o}}, {and} \bibinfo{person}{Michael Pradel}.}
  \bibinfo{year}{2022}\natexlab{}.
\newblock \showarticletitle{Meta Learning for Code Summarization}.
\newblock \bibinfo{journal}{\emph{{CoRR}}}  \bibinfo{volume}{2201.08310}
  (\bibinfo{year}{2022}), 5.
\newblock
\urldef\tempurl%
\url{https://doi.org/10.48550/arxiv.2201.08310}
\showDOI{\tempurl}


\bibitem[\protect\citeauthoryear{Reimers and Gurevych}{Reimers and
  Gurevych}{2019}]%
        {reimers-2019-sentence-bert}
\bibfield{author}{\bibinfo{person}{Nils Reimers} {and} \bibinfo{person}{Iryna
  Gurevych}.} \bibinfo{year}{2019}\natexlab{}.
\newblock \showarticletitle{Sentence-{BERT}: Sentence Embeddings using
  {S}iamese {BERT}-Networks}. In \bibinfo{booktitle}{\emph{Proceedings of the
  2019 Conference on Empirical Methods in Natural Language Processing and the
  9th International Joint Conference on Natural Language Processing}}.
  \bibinfo{publisher}{Association for Computational Linguistics},
  \bibinfo{address}{Hong Kong, China}, \bibinfo{pages}{3982--3992}.
\newblock
\urldef\tempurl%
\url{https://doi.org/10.18653/v1/D19-1410}
\showDOI{\tempurl}


\bibitem[\protect\citeauthoryear{Scherrer}{Scherrer}{2020}]%
        {scherrer-2020-tapaco}
\bibfield{author}{\bibinfo{person}{Yves Scherrer}.}
  \bibinfo{year}{2020}\natexlab{}.
\newblock \showarticletitle{{T}a{P}a{C}o: A Corpus of Sentential Paraphrases
  for 73 Languages}. In \bibinfo{booktitle}{\emph{Proceedings of the 12th
  Language Resources and Evaluation Conference}}. \bibinfo{publisher}{European
  Language Resources Association}, \bibinfo{address}{Marseille, France},
  \bibinfo{pages}{6868--6873}.
\newblock


\bibitem[\protect\citeauthoryear{Shi, Keneshloo, Ramakrishnan, and Reddy}{Shi
  et~al\mbox{.}}{2021}]%
        {Shi2021neural}
\bibfield{author}{\bibinfo{person}{Tian Shi}, \bibinfo{person}{Yaser
  Keneshloo}, \bibinfo{person}{Naren Ramakrishnan}, {and}
  \bibinfo{person}{Chandan~K Reddy}.} \bibinfo{year}{2021}\natexlab{}.
\newblock \showarticletitle{Neural abstractive text summarization with
  sequence-to-sequence models}.
\newblock \bibinfo{journal}{\emph{ACM Transactions on Data Science}}
  \bibinfo{volume}{2}, \bibinfo{number}{1} (\bibinfo{year}{2021}),
  \bibinfo{pages}{1--37}.
\newblock
\urldef\tempurl%
\url{https://doi.org/10.1145/3419106}
\showDOI{\tempurl}


\bibitem[\protect\citeauthoryear{Smilkov, Thorat, Nicholson, Reif, Vi{\'e}gas,
  and Wattenberg}{Smilkov et~al\mbox{.}}{2016}]%
        {smilkov2016embedding}
\bibfield{author}{\bibinfo{person}{Daniel Smilkov}, \bibinfo{person}{Nikhil
  Thorat}, \bibinfo{person}{Charles Nicholson}, \bibinfo{person}{Emily Reif},
  \bibinfo{person}{Fernanda~B Vi{\'e}gas}, {and} \bibinfo{person}{Martin
  Wattenberg}.} \bibinfo{year}{2016}\natexlab{}.
\newblock \showarticletitle{Embedding projector: Interactive visualization and
  interpretation of embeddings}.
\newblock \bibinfo{journal}{\emph{{CoRR}}}  \bibinfo{volume}{1611.05469}
  (\bibinfo{year}{2016}), 4.
\newblock
\urldef\tempurl%
\url{https://doi.org/10.48550/arxiv.1611.05469}
\showDOI{\tempurl}


\bibitem[\protect\citeauthoryear{Soo~Yi, Melton, Stasko, and Jacko}{Soo~Yi
  et~al\mbox{.}}{2005}]%
        {Soo2005dust}
\bibfield{author}{\bibinfo{person}{Ji Soo~Yi}, \bibinfo{person}{Rachel Melton},
  \bibinfo{person}{John Stasko}, {and} \bibinfo{person}{Julie~A Jacko}.}
  \bibinfo{year}{2005}\natexlab{}.
\newblock \showarticletitle{Dust \& magnet: multivariate information
  visualization using a magnet metaphor}.
\newblock \bibinfo{journal}{\emph{Information visualization}}
  \bibinfo{volume}{4}, \bibinfo{number}{4} (\bibinfo{year}{2005}),
  \bibinfo{pages}{239--256}.
\newblock
\urldef\tempurl%
\url{https://doi.org/10.1057/palgrave.ivs.9500099}
\showDOI{\tempurl}


\bibitem[\protect\citeauthoryear{Tan, Wan, and Xiao}{Tan et~al\mbox{.}}{2017}]%
        {Tan2017abstractive}
\bibfield{author}{\bibinfo{person}{Jiwei Tan}, \bibinfo{person}{Xiaojun Wan},
  {and} \bibinfo{person}{Jianguo Xiao}.} \bibinfo{year}{2017}\natexlab{}.
\newblock \showarticletitle{Abstractive document summarization with a
  graph-based attentional neural model}. In
  \bibinfo{booktitle}{\emph{Proceedings of the Annual Meeting of the
  Association for Computational Linguistics (Volume 1: Long Papers)}}.
  \bibinfo{publisher}{Association for Computational Linguistics},
  \bibinfo{address}{Online}, \bibinfo{pages}{1171--1181}.
\newblock


\bibitem[\protect\citeauthoryear{Tipping and Bishop}{Tipping and
  Bishop}{1999}]%
        {10.1162/089976699300016728}
\bibfield{author}{\bibinfo{person}{Michael~E. Tipping} {and}
  \bibinfo{person}{Christopher~M. Bishop}.} \bibinfo{year}{1999}\natexlab{}.
\newblock \showarticletitle{Mixtures of Probabilistic Principal Component
  Analyzers}.
\newblock \bibinfo{journal}{\emph{Neural Computation}} \bibinfo{volume}{11},
  \bibinfo{number}{2} (\bibinfo{date}{02} \bibinfo{year}{1999}),
  \bibinfo{pages}{443--482}.
\newblock
\urldef\tempurl%
\url{https://doi.org/10.1162/089976699300016728}
\showDOI{\tempurl}


\bibitem[\protect\citeauthoryear{van~der Maaten and Hinton}{van~der Maaten and
  Hinton}{2008}]%
        {JMLR:v9:vandermaaten08a}
\bibfield{author}{\bibinfo{person}{Laurens van~der Maaten} {and}
  \bibinfo{person}{Geoffrey Hinton}.} \bibinfo{year}{2008}\natexlab{}.
\newblock \showarticletitle{Visualizing Data using {t-SNE}}.
\newblock \bibinfo{journal}{\emph{Journal of Machine Learning Research}}
  \bibinfo{volume}{9}, \bibinfo{number}{86} (\bibinfo{year}{2008}),
  \bibinfo{pages}{2579--2605}.
\newblock


\bibitem[\protect\citeauthoryear{Vaswani, Shazeer, Parmar, Uszkoreit, Jones,
  Gomez, Kaiser, and Polosukhin}{Vaswani et~al\mbox{.}}{2017}]%
        {vaswani2017attention}
\bibfield{author}{\bibinfo{person}{Ashish Vaswani}, \bibinfo{person}{Noam
  Shazeer}, \bibinfo{person}{Niki Parmar}, \bibinfo{person}{Jakob Uszkoreit},
  \bibinfo{person}{Llion Jones}, \bibinfo{person}{Aidan~N Gomez},
  \bibinfo{person}{{\L}ukasz Kaiser}, {and} \bibinfo{person}{Illia
  Polosukhin}.} \bibinfo{year}{2017}\natexlab{}.
\newblock \showarticletitle{Attention is all you need}.
\newblock \bibinfo{journal}{\emph{Advances in neural information processing
  systems}}  \bibinfo{volume}{30} (\bibinfo{year}{2017}), 11.
\newblock


\bibitem[\protect\citeauthoryear{Wang, Singh, Michael, Hill, Levy, and
  Bowman}{Wang et~al\mbox{.}}{2018}]%
        {wang2018glue}
\bibfield{author}{\bibinfo{person}{Alex Wang}, \bibinfo{person}{Amanpreet
  Singh}, \bibinfo{person}{Julian Michael}, \bibinfo{person}{Felix Hill},
  \bibinfo{person}{Omer Levy}, {and} \bibinfo{person}{Samuel Bowman}.}
  \bibinfo{year}{2018}\natexlab{}.
\newblock \showarticletitle{{GLUE}: A Multi-Task Benchmark and Analysis
  Platform for Natural Language Understanding}. In
  \bibinfo{booktitle}{\emph{Proceedings of the {EMNLP} Workshop
  {B}lackbox{NLP}: Analyzing and Interpreting Neural Networks for {NLP}}}.
  \bibinfo{publisher}{Association for Computational Linguistics},
  \bibinfo{address}{Brussels, Belgium}, \bibinfo{pages}{353--355}.
\newblock
\urldef\tempurl%
\url{https://doi.org/10.18653/v1/W18-5446}
\showDOI{\tempurl}


\bibitem[\protect\citeauthoryear{Williams, Nangia, and Bowman}{Williams
  et~al\mbox{.}}{2018}]%
        {N18-1101}
\bibfield{author}{\bibinfo{person}{Adina Williams}, \bibinfo{person}{Nikita
  Nangia}, {and} \bibinfo{person}{Samuel Bowman}.}
  \bibinfo{year}{2018}\natexlab{}.
\newblock \showarticletitle{A Broad-Coverage Challenge Corpus for Sentence
  Understanding through Inference}. In \bibinfo{booktitle}{\emph{Proceedings of
  the Conference of the North American Chapter of the Association for
  Computational Linguistics: Human Language Technologies, Volume 1 (Long
  Papers)}}. \bibinfo{publisher}{Association for Computational Linguistics},
  \bibinfo{address}{New Orleans, Louisiana}, \bibinfo{pages}{1112--1122}.
\newblock


\bibitem[\protect\citeauthoryear{Wu, Ouyang, Ziegler, Stiennon, Lowe, Leike,
  and Christiano}{Wu et~al\mbox{.}}{2021}]%
        {Wu2021recursively}
\bibfield{author}{\bibinfo{person}{Jeff Wu}, \bibinfo{person}{Long Ouyang},
  \bibinfo{person}{Daniel~M Ziegler}, \bibinfo{person}{Nissan Stiennon},
  \bibinfo{person}{Ryan Lowe}, \bibinfo{person}{Jan Leike}, {and}
  \bibinfo{person}{Paul Christiano}.} \bibinfo{year}{2021}\natexlab{}.
\newblock \showarticletitle{Recursively Summarizing Books with Human Feedback}.
\newblock \bibinfo{journal}{\emph{{CoRR}}}  \bibinfo{volume}{2109.10862}
  (\bibinfo{year}{2021}), 37.
\newblock
\urldef\tempurl%
\url{https://doi.org/10.48550/arXiv.2109.10862}
\showDOI{\tempurl}


\bibitem[\protect\citeauthoryear{Yang, Qu, Shen, Liu, Zhao, and Zhu}{Yang
  et~al\mbox{.}}{2018}]%
        {yang-etal-2018-aspect}
\bibfield{author}{\bibinfo{person}{Min Yang}, \bibinfo{person}{Qiang Qu},
  \bibinfo{person}{Ying Shen}, \bibinfo{person}{Qiao Liu}, \bibinfo{person}{Wei
  Zhao}, {and} \bibinfo{person}{Jia Zhu}.} \bibinfo{year}{2018}\natexlab{}.
\newblock \showarticletitle{Aspect and Sentiment Aware Abstractive Review
  Summarization}. In \bibinfo{booktitle}{\emph{Proceedings of the International
  Conference on Computational Linguistics}}. \bibinfo{publisher}{Association
  for Computational Linguistics}, \bibinfo{address}{Santa Fe, New Mexico, USA},
  \bibinfo{pages}{1110--1120}.
\newblock


\bibitem[\protect\citeauthoryear{Zhang, Zhao, Saleh, and Liu}{Zhang
  et~al\mbox{.}}{2020b}]%
        {Zhang2020pegasus}
\bibfield{author}{\bibinfo{person}{Jingqing Zhang}, \bibinfo{person}{Yao Zhao},
  \bibinfo{person}{Mohammad Saleh}, {and} \bibinfo{person}{Peter Liu}.}
  \bibinfo{year}{2020}\natexlab{b}.
\newblock \showarticletitle{Pegasus: Pre-training with extracted gap-sentences
  for abstractive summarization}. In \bibinfo{booktitle}{\emph{International
  Conference on Machine Learning}}. \bibinfo{publisher}{PMLR},
  \bibinfo{address}{online}, \bibinfo{pages}{11328--11339}.
\newblock


\bibitem[\protect\citeauthoryear{Zhang, Kishore, Wu, Weinberger, and
  Artzi}{Zhang et~al\mbox{.}}{2020a}]%
        {zhang2020bertscore}
\bibfield{author}{\bibinfo{person}{Tianyi Zhang}, \bibinfo{person}{Varsha
  Kishore}, \bibinfo{person}{Felix Wu}, \bibinfo{person}{Kilian~Q. Weinberger},
  {and} \bibinfo{person}{Yoav Artzi}.} \bibinfo{year}{2020}\natexlab{a}.
\newblock \showarticletitle{{BERTScore:} Evaluating Text Generation with
  {BERT}}. In \bibinfo{booktitle}{\emph{Proceedings of the International
  Conference on Learning Representations}}.
  \bibinfo{publisher}{Openreview.net}, \bibinfo{address}{online}, 43.
\newblock


\bibitem[\protect\citeauthoryear{Zhang, Chandrasegaran, and Ma}{Zhang
  et~al\mbox{.}}{2021}]%
        {zhang2021conceptscope}
\bibfield{author}{\bibinfo{person}{Xiaoyu Zhang}, \bibinfo{person}{Senthil
  Chandrasegaran}, {and} \bibinfo{person}{Kwan-Liu Ma}.}
  \bibinfo{year}{2021}\natexlab{}.
\newblock \showarticletitle{{ConceptScope:} Organizing and visualizing
  knowledge in documents based on domain ontology}. In
  \bibinfo{booktitle}{\emph{Proceedings of the ACM CHI conference on human
  factors in computing systems}}. \bibinfo{publisher}{ACM},
  \bibinfo{address}{New York, NY}, \bibinfo{pages}{19:1--19:13}.
\newblock
\urldef\tempurl%
\url{https://doi.org/10.1145/3411764.3445396}
\showDOI{\tempurl}


\bibitem[\protect\citeauthoryear{Zhang, Baldridge, and He}{Zhang
  et~al\mbox{.}}{2019}]%
        {paws2019naacl}
\bibfield{author}{\bibinfo{person}{Yuan Zhang}, \bibinfo{person}{Jason
  Baldridge}, {and} \bibinfo{person}{Luheng He}.}
  \bibinfo{year}{2019}\natexlab{}.
\newblock \showarticletitle{{PAWS}: Paraphrase Adversaries from Word
  Scrambling}. In \bibinfo{booktitle}{\emph{Proceedings of the Conference of
  the North {A}merican Chapter of the Association for Computational
  Linguistics: Human Language Technologies, Volume 1 (Long and Short Papers)}}.
  \bibinfo{publisher}{Association for Computational Linguistics},
  \bibinfo{address}{Minneapolis, Minnesota}, \bibinfo{pages}{1298--1308}.
\newblock
\urldef\tempurl%
\url{https://doi.org/10.18653/v1/N19-1131}
\showDOI{\tempurl}


\end{thebibliography}
